\documentclass{aastex63}
\usepackage{float}
\usepackage[hyphens]{url}  
\usepackage{hyperref}

\newcommand{\Nspectra}{59,266 }
\newcommand{\Nstars}{24,130 }
\newcommand{\Ntraining}{1,675 }
\newcommand{\Nfits}{54,729 }

\newcommand{\teff}{$T_{eff}$}
\newcommand{\logg}{$\log g$}
\newcommand{\feh}{$[Fe/H]$}
\newcommand{\alphafe}{$[\alpha/Fe]$}
\newcommand{\vmicro}{$\log(v_{micro})$}

\received{August 30, 2021 }
\revised{November 8, 2021}
\accepted{November 20, 2021 }
\submitjournal{\textit{The Astronomical Journal}}

\shorttitle{SDSS-IV MaStar: Data-driven Parameter Derivation for the MaStar Stellar Library}
\shortauthors{Imig et al. 2021}
\graphicspath{{./}{figures/}}

\begin{document}

\title{SDSS-IV MaStar: Data-driven Parameter Derivation for the MaStar Stellar Library}

\author[0000-0003-2025-3585]{Julie Imig}
\affiliation{Department of Astronomy, New Mexico State University, P.O.Box 30001, MSC 4500, Las Cruces, NM, 88033, USA}

\author[0000-0002-9771-9622]{Jon A. Holtzman}
\affiliation{Department of Astronomy, New Mexico State University, P.O.Box 30001, MSC 4500, Las Cruces, NM, 88033, USA}

\author[0000-0003-1025-1711]{Renbin Yan}
\affiliation{Department of Physics and Astronomy, University of Kentucky, 505 Rose St., Lexington, KY 40506-0057, USA}

\author{Daniel Lazarz}
\affiliation{Department of Physics and Astronomy, University of Kentucky, 505 Rose St., Lexington, KY 40506-0057, USA}

\author[0000-0001-8821-0309]{Yanping Chen}
\affiliation{New York University Abu Dhabi, Abu Dhabi, P.O. Box 129188, United Arab Emirates}

\author{Lewis Hill}
\affiliation{Institute of Cosmology \& Gravitation, University of Portsmouth, Dennis Sciama Building, Portsmouth, PO1 3FX, UK}

\author[0000-0002-6325-5671]{Daniel Thomas}
\affiliation{Institute of Cosmology \& Gravitation, University of Portsmouth, Dennis Sciama Building, Portsmouth, PO1 3FX, UK}

\author[0000-0001-7711-3677]{Claudia Maraston}
\affiliation{Institute of Cosmology \& Gravitation, University of Portsmouth, Dennis Sciama Building, Portsmouth, PO1 3FX, UK}

\author[0000-0001-8302-0565]{Moire M. K. Prescott}
\affiliation{Department of Astronomy, New Mexico State University, P.O.Box 30001, MSC 4500, Las Cruces, NM, 88033, USA}

\author[0000-0003-1479-3059]{Guy S. Stringfellow}
\affiliation{Center for Astrophysics and Space Astronomy, Department of Astrophysical and Planetary Sciences, University of Colorado, 389 UCB, Boulder, CO 80309-0389, USA}

\author[0000-0002-3601-133X]{Dmitry Bizyaev}
\affiliation{Apache Point Observatory and New Mexico State University, P.O. Box 59, Sunspot, NM, 88349-0059, USA}
\affiliation{Sternberg Astronomical Institute, Moscow State University, Moscow}

\author[0000-0002-1691-8217]{Rachael L. Beaton}
\affiliation{The Observatories of the Carnegie Institution for Science, 813 Santa Barbara Street, Pasadena, CA 91101, USA}

\author{Niv Drory}
\affiliation{McDonald Observatory, The University of Texas at Austin, 1 University Station, Austin, TX 78712, USA}




\begin{abstract}

The MaNGA Stellar Library (MaStar) is a large collection of high-quality empirical stellar spectra designed to cover all spectral types and ideal for use in the stellar population analysis of galaxies observed in the Mapping Nearby Galaxies at Apache Point Observatory (MaNGA) survey. The library contains \Nspectra spectra of \Nstars unique stars with spectral resolution $R\sim1800$ and covering a wavelength range of $3,622-10,354$ \AA. In this work, we derive five physical parameters for each spectrum in the library: effective temperature (\teff), surface gravity (\logg), metalicity (\feh), micro-turbulent velocity (\vmicro), and alpha-element abundance (\alphafe). These parameters are derived with a flexible data-driven algorithm that uses a neural network model. We train a neural network using the subset of \Ntraining MaStar targets that have also been observed in the Apache Point Observatory Galactic Evolution Experiment (APOGEE), adopting the independently-derived APOGEE Stellar Parameter and Chemical Abundance Pipeline (ASPCAP) parameters for this reference set. For the regions of parameter space not well represented by the APOGEE training set ($7,000 \leq T \leq 30,000$ K), we supplement with theoretical model spectra. We present our derived parameters along with an analysis of the uncertainties and comparisons to other analyses from the literature.

\end{abstract}


\keywords{Spectroscopy – Fundamental parameters of stars – Stellar abundances – Astronomy data analysis – Astrostatistics – Surveys}


\section{Introduction}
\label{section:Introduction}
 Stellar spectral libraries play an important role in understanding a wide range of stellar, Galactic, and extragalactic astrophysics. In stellar and Galactic astronomy, stellar spectral libraries can be used to estimate stellar parameters, infer interstellar extinction or distances, or model stellar continua. For extragalactic astronomy, stellar libraries are used to fit an integrated galaxy spectrum with stellar population synthesis modeling and derive star formation histories, initial mass functions, and observed redshifts \citep[e.g.,][]{Tinsley1972,Bruzual1983,Fioc1997,Leitherer_1999,Bruzual_2003,Cardiel2003,Thomas2005,Maraston_2005,Coelho2007,Conroy_2009,Vazdekis_2010, Bruzual2011,Maraston_2011, Conroy_2013,Goddard2016,R_ck_2016,Vazdekis2016,GarciaBenito2017,maraston2020}. Our understanding of galaxy evolution has greatly expanded over the past few years by applying these methods to integral field unit surveys that have observed thousands of spatially-resolved galaxies, e.g., the Mapping Nearby Galaxies at Apache Point Observatory \citep[MaNGA;][]{MaNGAOverview}, the Calar Alto Legacy Integral Field Area survey \citep[CALIFA;][]{CALIFA}, and the Sydney-Australian-Astronomical-Observatory Multi-object Integral-Field Spectrograph galaxy survey \citep[SAMI;][]{Croom2021}.

Stellar spectral libraries can be composed of theoretical or empirical spectra, each approach having strengths and weaknesses. Theoretical stellar spectral libraries \citep[e.g.,][]{Kurucz1979,Diaz1989, Lejeune1997,Westera2002,atlas9,Decin2004,MARCS2008,apogeegrid,BOSZ,Allende_Prieto_2018} are calculated using models of radiative-transfer processes through a stellar atmosphere, and are powerful in their ability to provide spectra for any combination of stellar parameters at any resolution. However, they can be limited by physical effects that are difficult to model, such as non-local-thermodynamic-equilibrium (non-LTE) effects, spherical geometry, line-blanketing, atmospheric expansion, and non-radiative heating. Absorption features can also be missing from theoretical spectra altogether, due to incomplete line lists for atomic or molecular species where precise atomic data is not available for all transitions. As a result, theoretical models cannot yet accurately reproduce the spectra for some stars \citep[e.g.,][]{Kurucz1979,Kurucz2011,Dupree_2016}. Empirical stellar spectral libraries \citep[e.g.,][]{Gunn1983,Pickles1985,Silva1992,Pickles1998,Cenarro2001,Prugniel2001,LeBorgne2003,Valdes2004,gregg2006,MILES2006,Kirby2011,Chen2014,mastar2019}, on the other hand, are obtained by observing real stars, avoiding many of these concerns. At the same time, empirical libraries are limited by the wavelength range and spectral resolution of the observing instrument, as well as the range of parameter space that they can cover; clearly, spectra cannot be obtained for types of stars that do not exist within an observable distance.

The MaNGA Stellar Library \citep[MaStar;][]{mastar2019} is a large, well-calibrated, and high quality empirical collection of stellar spectra released as part of the Sloan Digital Sky Survey \citep[SDSS-IV;][]{sdss4}. A wide variety of stars were intentionally targeted by the MaStar project, providing extensive coverage in effective temperature (T$_{eff}$), surface gravity (log$g$), iron metallicity ($[Fe/H]$), and alpha-element abundance ($[\alpha/Fe]$). Alpha-element abundance estimates are of particular interest for modeling galaxies, as the $[\alpha/Fe]$ ratio is known to be informative of a galaxy's enrichment history and traces the timescales over which star formation occurs \citep[e.g.,][]{mb1990, Thomas2005}. The MaStar catalog will be particularly useful for the stellar population synthesis analysis of galaxies observed in the MaNGA survey. Both the MaStar stellar library and the MaNGA survey were obtained using same instrument \citep{Drory2015}, ensuring that the galactic spectra from MaNGA and the stellar spectra from MaStar cover the same wavelength range with the same resolution, making the MaStar library ideal for analyzing these galaxies \citep[e.g., MaStar-based stellar population models from][]{maraston2020}.

Before an empirical stellar library can be used for any application, a crucial first step is to consistently and accurately determine the physical parameters associated with each spectrum. For use in stellar or Galactic astronomy, the library can be used to pick out a star of a particular spectral type or evolutionary phase, in which case the parameters are necessary to identify the desired spectrum for comparison. For applications involving stellar population synthesis of galaxies, stellar libraries can be used to construct a spectrum of a simple stellar population, a family of stars with a single age and metallicity. In this case, the parameters of the spectral library are used to associate a single point on an isochrone to a corresponding empirical spectrum in the stellar library \citep[e.g., ][]{Bruzual_2003,Coelho2007,Percival2008,Vazdekis_2010,maraston2020}. Constructing the wide variety of simple stellar population models needed to realistically model observed galaxies requires a spectral library that includes stars from a wide range of parameter space. 

In this paper, we present a semi-empirical approach for deriving the the effective temperature, T$_{eff}$, surface gravity, log$g$, metallicity, $[Fe/H]$, microturbulent velocity, \vmicro, and alpha-element abundance, [$\alpha$/$Fe$], for \Nfits spectra in the MaStar stellar library. The data-driven aspect of this methodology utilizes a subset of the MaStar library with independently determined parameters as a reference set, in a sense, using the MaStar library to self-consistently estimate the parameters for the MaStar library.  The parameters for our empirical training set are adopted from the APOGEE Stellar Parameters and Chemical Abundance Pipeline \citep[ASPCAP;][]{aspcap, aspcap2018, aspcapdr17} catalog. This empirical training set is then combined with theoretical spectra from \cite{Allende_Prieto_2018} to extend the training set to hotter temperatures. We train a neural network on this set to reproduce a spectrum as a function of its parameters, then use the network to determine the parameters of the remaining spectra in the library. This work, \cite{Chen2021}, \cite{lazarz2021}, and \cite{hill2021} each present an alternative set of stellar parameters for the MaStar Stellar Library determined using different methods. All four parameter catalogs will be published in \cite{mastar2021}, along with a detailed comparison between the four methodologies and the overview of the final data release of MaStar.

A description of the MaStar data and the subsets selected as a reference sample can be found in Section \ref{section:Data}. We motivate our methodology and present the architecture and performance of the neural network in Section \ref{section:Methods}. The derived parameter results are presented and validated with further comparisons in Section \ref{section:Results}. Section \ref{section:Discussion} contains a discussion of potential systematics in our results. Finally, we conclude in Section \ref{section:Conclusions}.

\section{Data}
\label{section:Data}

\subsection{MaStar Overview}
\label{section:Datamastar}

The first release of the MaStar Stellar library is presented in \cite{mastar2019}, and the final version of the library will be detailed in \cite{mastar2021}. In brief, the library contains \Nspectra individual spectra (visits) of \Nstars unique stars, covering the wavelength range of $3,622 - 10,354$ \AA at resolution $R \sim 1800$. The stellar spectra were obtained using the Baryon Oscillation Spectroscopic Survey (BOSS) spectrograph \citep{bossspectrograph, Drory2015} mounted on the Apache Point Observatory 2.5m telescope \citep{apo25m}, using the same fiber bundles that the MaNGA survey used to observe spatially-resolved nearby galaxies \citep{Drory2015,MaNGAOverview}. The spectra were reduced using the MPL-11 version of the MaNGA Data Reduction Pipeline \citep[DRP;][]{mangadrp}. The flux calibration of the library is accurate to 4\%, and the majority (90\%) of spectra in the library have $S/N \geq 50$ \citep{mastar2021}.

Among the \Nstars unique stars in the MaStar library, about half of them (12,345) were targeted as standard stars used for flux calibration. There are an additional 11,817 unique science target stars in MaStar, with 32 stars being both standard and science targets. Science targets for the MaStar stellar library were selected using a targeting algorithm designed to obtain spectra representing as wide a range of parameter space as possible \citep[Appendix D]{mastar2019}. Stellar parameters from existing catalogs were used to select potential targets: (\teff, \logg, \feh) were adopted from LAMOST \citep{LAMOST_Luo2015} and SEGUE \citep{Lee_2008}, and (\teff, \logg, \feh, \alphafe) from APOGEE \citep{aspcap}. In building the library, higher targeting priority was placed on stars from underpopulated regions of parameter space when compared to the already-observed fraction of MaStar, effectively placing higher weights on “rare" stars. As a result, the MaStar library targeted a broad range of expected stellar parameters:

\begin{itemize}
    \item $3000$ K $\lessapprox T_{eff} \lessapprox 30,000$ K
    \item $-0.5$ dex $\lessapprox \log g \lessapprox 5.0$ dex
    \item $-2.5$ dex $\lessapprox [Fe/H] \lessapprox 0.5$ dex
    \item $-0.2$ dex $\lessapprox [\alpha/Fe] \lessapprox 0.6$ dex
\end{itemize}

\subsection{MaStar Spectra}
\label{section:DataSpectra}

Here, we describe the steps taken to prepare the MaStar data for our procedure, and highlight a number of other factors that might influence the quality of our results.

In the MaStar library, a number of different quality flags have been defined to document potentially problematic spectra, as described in \cite{mastar2019} and updated in \cite{mastar2021}. In this paper, we work with the “good visits" version of the library, containing only high quality spectra on an individual visit basis. Spectra that suffer from bad sky subtraction, scattered light, uncertain radial velocity measurements, issues manually identified from visual inspection, non-stellar targets, and catastrophically low signal-to-noise ratios ($S/N \leq 15$) were flagged and removed from this version of the library.

\begin{figure}
    \centering
    \includegraphics[width=\textwidth]{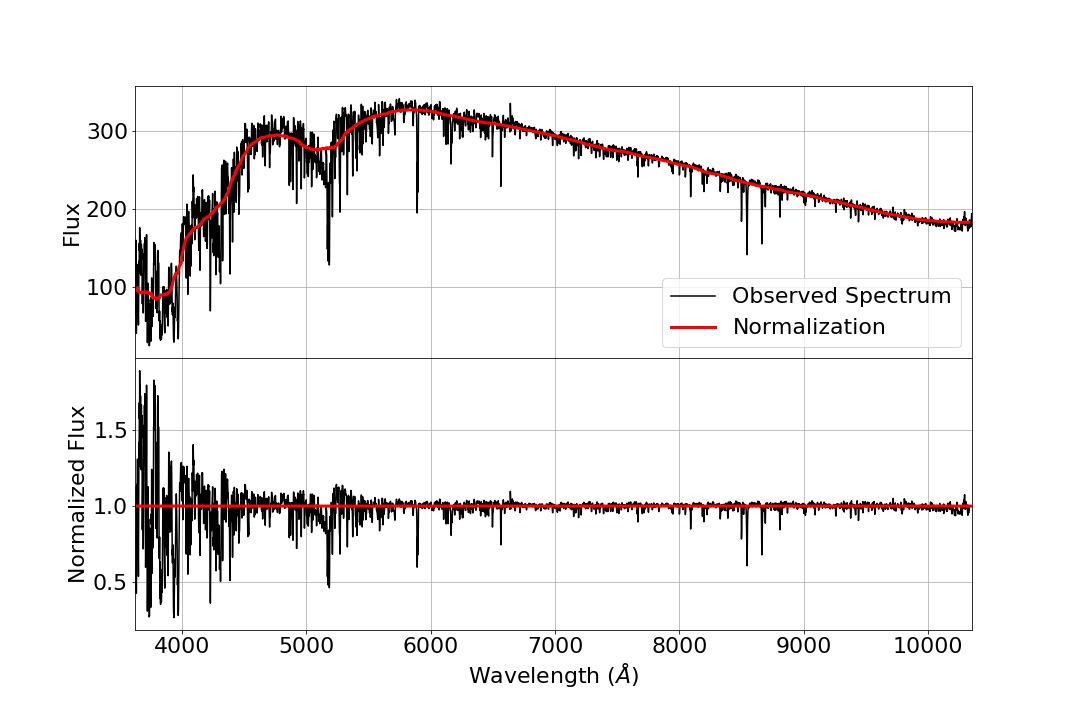}
    \caption{An example of the pseudo-continuum normalization procedure described in section \ref{section:DataSpectra}. Top: An unnormalized spectrum (black line) and a smoothed version of the spectrum created with a 400-pixel window running median (red line), which is the estimate of the pseudo-continuum. Bottom: The resulting normalized spectrum, made by dividing out the pseudo-continuum from the original spectrum. 
}
    \label{fig:norm}
\end{figure}

The spectra in MaStar are not continuum-corrected for interstellar reddening. In order to minimize continuum-shape-induced effects that have the potential to influence any derived parameters, such as imperfect flux calibration and interstellar reddening, we normalize each spectrum in the library employing a running-median scheme, dividing each pixel by the median value of the spectrum within a 400 pixel wide window. The window was selected to be wide enough to preserve large absorption features in the spectra, such as the TiO bands in cool stars. An example of the normalization process is shown in Figure \ref{fig:norm}. Since the median is used to approximate the continuum, this is a “psuedo-continuum" normalization scheme and not a true continuum identification. 

The resolution of MaStar varies across wavelengths and differs from fiber-to-fiber. Multiple spectra of the same star, if taken on different nights, might not have the same resolution as a function of wavelength. As a result, MaStar spectra are presented on an individual visit basis, and are not stacked together to create one spectrum for each unique star. We exploit this fact in section \ref{section:uncertainties} to test the consistency of our results, but our methodology uses the (incorrect) assumption that every variation in a spectrum is explained solely by its parameters; we do not take into consideration the variation in resolution between spectra. We explore the effects that this may have on our parameter results in Section \ref{section:Discussion:LSF}.

\subsection{APOGEE-MaStar Overlap Sample}
\label{section:Dataapogee}

The Apache Point Observatory Galactic Evolution Experiment \citep[APOGEE;][]{apogeeoverview} is a high-resolution ($R\sim20,000$) infrared spectroscopic survey that is also part of SDSS-IV \citep{sdss4}. The DR17 release of MaStar and APOGEE have observed 2,304 stars in common. This overlap sample provides an ideal reference set for determining the parameters of the MaStar library, as APOGEE provides precise, independently derived parameters for these stars. Specifically, we adopt the effective temperature, T$_{eff}$, surface gravity, log$g$, iron metallicity, $[Fe/H]$, microturbulent velocity, \vmicro, and alpha-element abundance, [$\alpha$/$Fe$], values from the APOGEE Stellar Parameter and Chemical Abundance Pipeline (ASPCAP) for use in our training set. 

The ASPCAP pipeline is described in detail in \cite{aspcap}, \cite{aspcap2018}, and \cite{aspcapdr17}, but a summary is provided here. First, ASPCAP compares the observed APOGEE spectra to a theoretical spectral library to determine atmospheric parameters ($T_{eff}, \log g$, \vmicro, \feh). The theoretical library was generated specifically for APOGEE using the MARCS model atmospheres \citep{apogeegrid}. The comparison is carried out using a multi-dimensional $\chi^2$ minimization by the code \verb|FERRE| \citep{ferre}. The second step determines chemical abundances for one element at a time by fitting confined windows of the spectrum, using the atmospheric parameters derived before while varying individual abundances. For our training set, we adopt the calibrated values of effective temperatures and surface gravities from ASPCAP. Effective temperatures are calibrated with zero-point offsets based on photometric effective temperatures for stars with low reddening, using the relations of \citet{GHB2009}. The surface gravity calibration is based on asteroseismic measurements for evolved stars (from a pre-release of the APOKASC-3 catalog, Pinnsoneault et al 2021, 
in preparation) and isochrone gravities for main sequence stars \citep{Berger2020}. Most of the calibrator stars fall within the range 4000 $<$ \teff $<$ 7000 and \logg $>$1. The metallicity (\feh) is uncalibrated, but atomic line data used to construct the synthetic spectral libraries that were fit to the data were adjusted to match high resolution spectra of the Sun and Arcturus \citep{Smith2021}.
Additional details are provided in \cite{aspcapdr17}.

The typical precision associated with ASPCAP parameters, based on repeat measurements of the same stars, is high. For the APOGEE-MaStar Overlap sample, the median ASPCAP uncertainties for stellar effective temperatures are reported within 1\%, surface gravities within 0.03 dex, metallicities and alpha-element abundances within 0.01 dex.

In addition to precision, the systematic inaccuracies that may be present in ASPCAP are important to understand for our work, as any inaccuracies in ASPCAP will be propagated through the neural network into our final results. A detailed study of the accuracy of the stellar parameters derived from ASPCAP can be found in \cite{aspcap2018}, where parameter results are compared to a high-quality sample of Gaia standard stars. They find that the calibrated effective temperatures and surface gravities are typically accurate within 100 K and 0.05 dex respectively. The accuracy of element abundances was assessed through a comparison to published high-resolution optical studies \citep{Brewer2016, daSilva2015,Jonsson2017,Lomaeva2019,Forseberg2019} of a subset of APOGEE stars. They found that ASPCAP metallicity estimates are typically accurate for metal-rich stars, but may have systematic offsets at the 0.05-0.1 dex level for metal-poor stars. Alpha-element abundances are among the most precise and the most accurate of all of the elements measured in ASPCAP, with a typical offset of less than 0.01 dex.

\begin{figure}
    \centering
    \includegraphics[width=\textwidth]{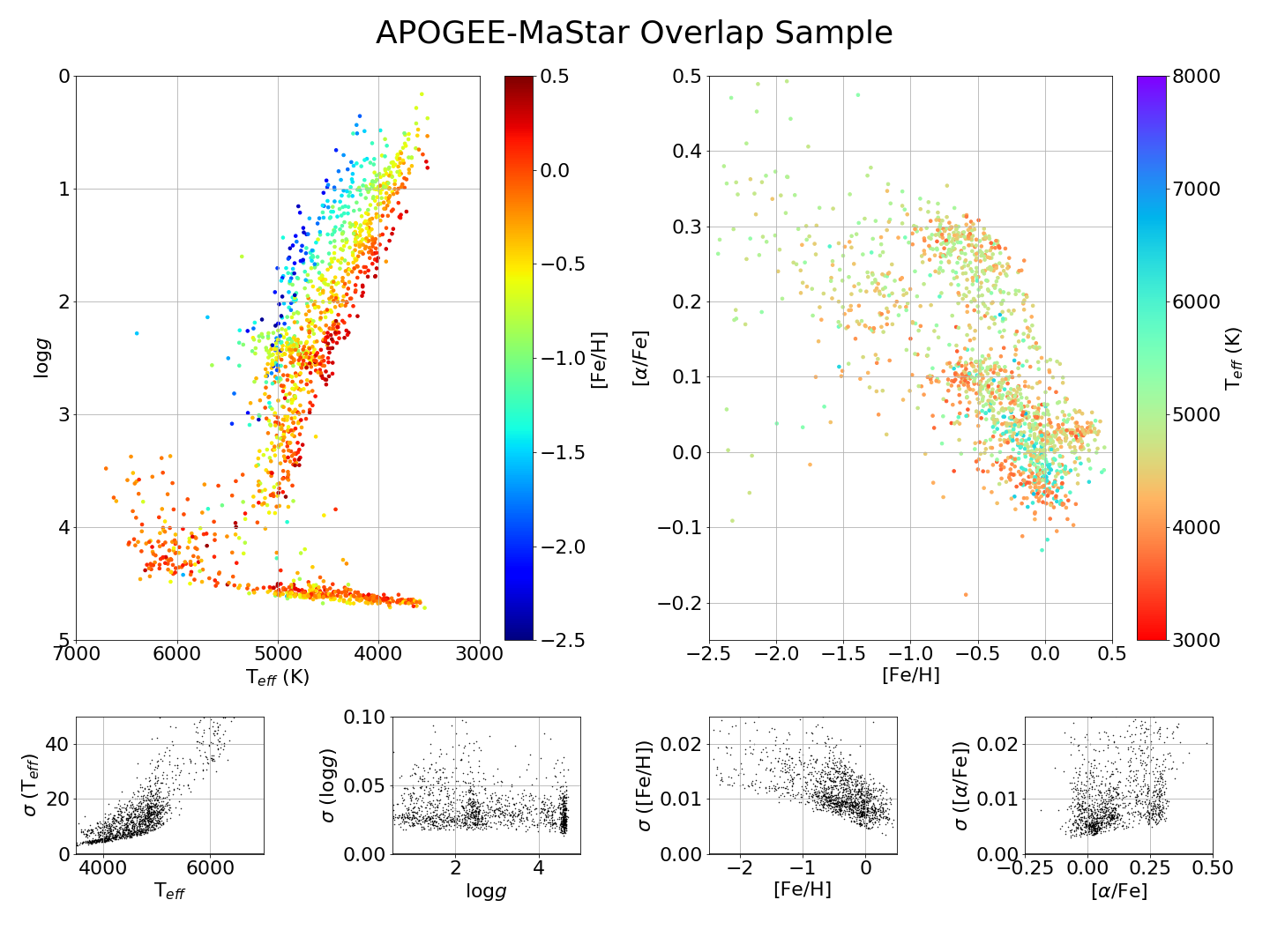}
    \caption{ASPCAP Parameter distribution of the APOGEE-MaStar overlap sample in a Kiel diagram (left) and the [$\alpha$/Fe]-[Fe/H] plane (right). Bottom Row: Reported precision} in the ASPCAP parameters for the APOGEE-MaStar overlap sample.
    \label{fig:APOGEETrainingCMD}
\end{figure}

Starting with the overlap sample, we removed potentially problematic stars, including stars identified as spectroscopic binaries or containing parameter quality flags in ASPCAP. Spectra with emission lines identified by MaStar bitmasks were also thrown out. The remaining empirical reference set is the concatenation of ASPCAP parameters and the corresponding MaStar spectra for \Ntraining unique stars. Multiple spectra of the same star were stacked across visits to reduce noise, coadding each spectrum after weighting by the inverse variance for each pixel. This was done to reduce noise in our reference set and ensure that the neural network receives the highest quality spectra as possible for training. The ASPCAP parameter distribution and precisions for this empirical reference set is shown in Figure \ref{fig:APOGEETrainingCMD}.

We note that this distribution of stars is not a perfect training set. Namely, in this sample there are very few (27) metal-poor dwarfs (\teff $\leq 6000$ K, \logg $ \geq 4.0$, \feh $ \leq -1.5$,) and a small number (22) of cool dwarfs overall (\teff $\leq 3750$ K, \logg $ \geq 4.0$). Neural networks can only be expected to perform well over regions where the training set is adequately populated. The lack of representation for these stars may introduce artifacts in our final parameter results. We return to this in Section \ref{results:QC}.

\subsection{Theoretical Reference Set}
\label{section:DataSynth}

APOGEE preferentially targets cooler stars that have more flux within APOGEE's infrared wavelength range; the hottest star in the APOGEE-MaStar Overlap sample is $T_{eff} \approx 7000$ K. However, the MaStar library includes targets much hotter than this, so we must extend our reference set to adequately represent the full range of temperatures present in the MaStar library. Therefore, in addition to the empirical reference set, we adopt a set of theoretical spectra from \cite{Allende_Prieto_2018} to extend our reference set to cover hotter temperatures ($6000 \leq T_{eff} \leq 30,000$ K) than those available from APOGEE. These synthetic spectra were computed using the plane-parallel ATLAS9 model atmospheres \citep{atlas9} and the radiative transfer code \verb|ASS|$\epsilon$\verb|T| \citep{koesterke2009}.  This library reproduces the spectra of real stars within 5\% for temperatures of $6000 \leq $\teff $\leq$ 20,000 K \citep[][Section 4]{Allende_Prieto_2018}, discussed further in Section \ref{section:MethodsMotivation}.

The models were convolved to match the median resolution of the MaStar library as a function of wavelength \citep[Figure 10]{mastar2019}. We choose to add \Ntraining synthetic spectra to the training set, the same number as empirical spectra in the APOGEE-MaStar Overlap sample. This ensures that the neural network receives an equal amount of information from both synthetic and empirical spectra, across hot and cool stars, and the resulting neural network model is not weighted towards one or the other. Additionally, there is a small amount of overlap between the synthetic and empirical reference sets; there are 89 stars in the empirical reference set with $T_{eff} \geq 6000$ K. 


Rather than adopt the grid as is, we choose to interpolate the model grid to produce a more physically-motivated distribution of stellar spectra. We randomly select points in $T_{eff}$, $\log g$, and $[Fe/H]$ from a set of \verb|PARSEC| stellar isochrones \citep{Bressan_2012} with temperature greater than 6000 K. For each selected isochrone point, we generate a random value for alpha-element abundance and microturbulence within the limits of the synthetic spectral grid. This does not reflect a physically-motivated distribution in these two parameters, but varying these has a much smaller effect on a spectrum than the former three parameters; there is no disadvantage to supplying a wider range of information to the neural network. The model spectrum corresponding to the resulting randomly-drawn parameter distribution is computed using a cubic Bezier interpolation with the code \verb|FERRE| in pure interpolation mode \citep{ferre}.

Although this theoretical reference set represents half of our total training set, stars in this temperature range ($T_{eff} \geq 7,000$ K) represent a minority (8\%) of the MaStar library as a whole. Extremely hot stars ($T_{eff} \geq 20,000$ K), where non-LTE effects may be more prevalent, are only 0.4\% of the MaStar library. The vast majority of MaStar targets will be covered within the range of the empirical training set. Any spectral inaccuracies in the synthetic spectra will only influence a small fraction of our results. However, hot stars are important in stellar population synthesis models, as they often dominate the light in some populations.

\begin{figure}
    \centering
    \includegraphics[width=\textwidth]{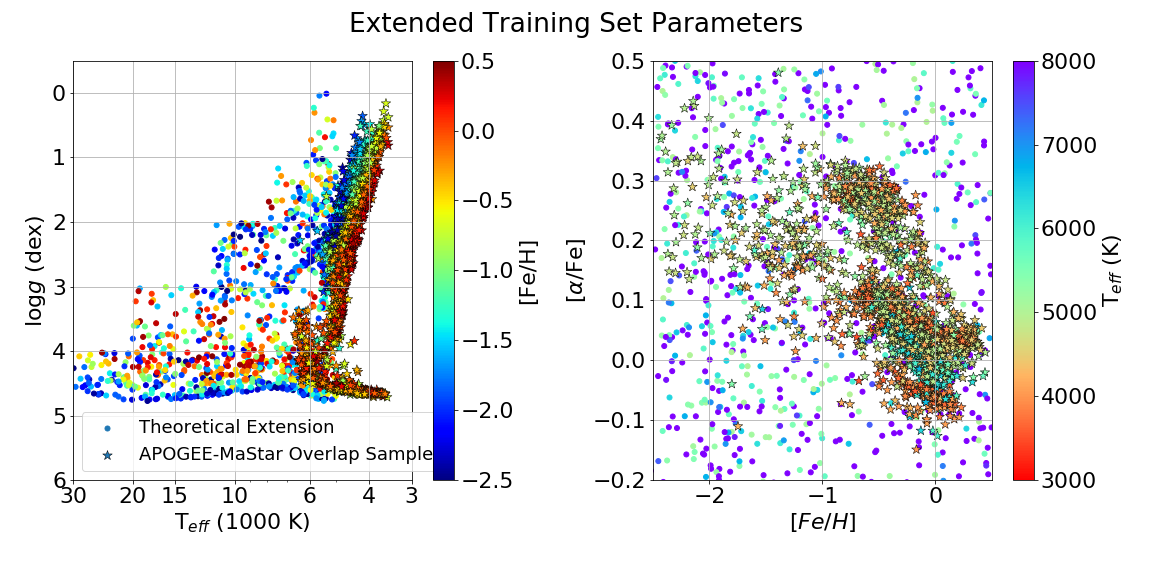}
    \caption{The full reference set which we use to train the neural network, shown in a Kiel diagram (Left) and the $[\alpha/Fe]-[Fe/H]$ plane (right). The training set includes parameters from ASPCAP (star points), and extends to hotter temperatures through the addition of randomly-drawn synthetic spectra (round points).}
    \label{fig:FullTrainingCMD}
\end{figure}

The parameter distribution for the full reference set (the empirical APOGEE-MaStar Overlap Sample and the theoretical extension) is shown in Figure  \ref{fig:FullTrainingCMD}.


\section{Methods}
\label{section:Methods}

\subsection{Motivation for a Data-Driven Approach}
\label{section:MethodsMotivation}

Our decision to adopt a data-driven approach and train our neural network partially using a sample of real stars is linked to the motivation for compiling an empirical stellar library like MaStar. While synthetic spectra made from theoretic model atmospheres can accurately reproduce observations of many stars, they cannot yet adequately reproduce the spectra of cooler dwarfs \citep[e.g.,][]{Kurucz1979,Dupree_2016}. The atmospheres of these stars contain an abundance of molecules such as water, methane, and titanium oxide, with transitions not well understood or not included in many modern linelists. The atmospheres of cooler dwarfs can also contain non-LTE effects which are often difficult to model.

\begin{figure}
    \centering
    \includegraphics[width=\textwidth]{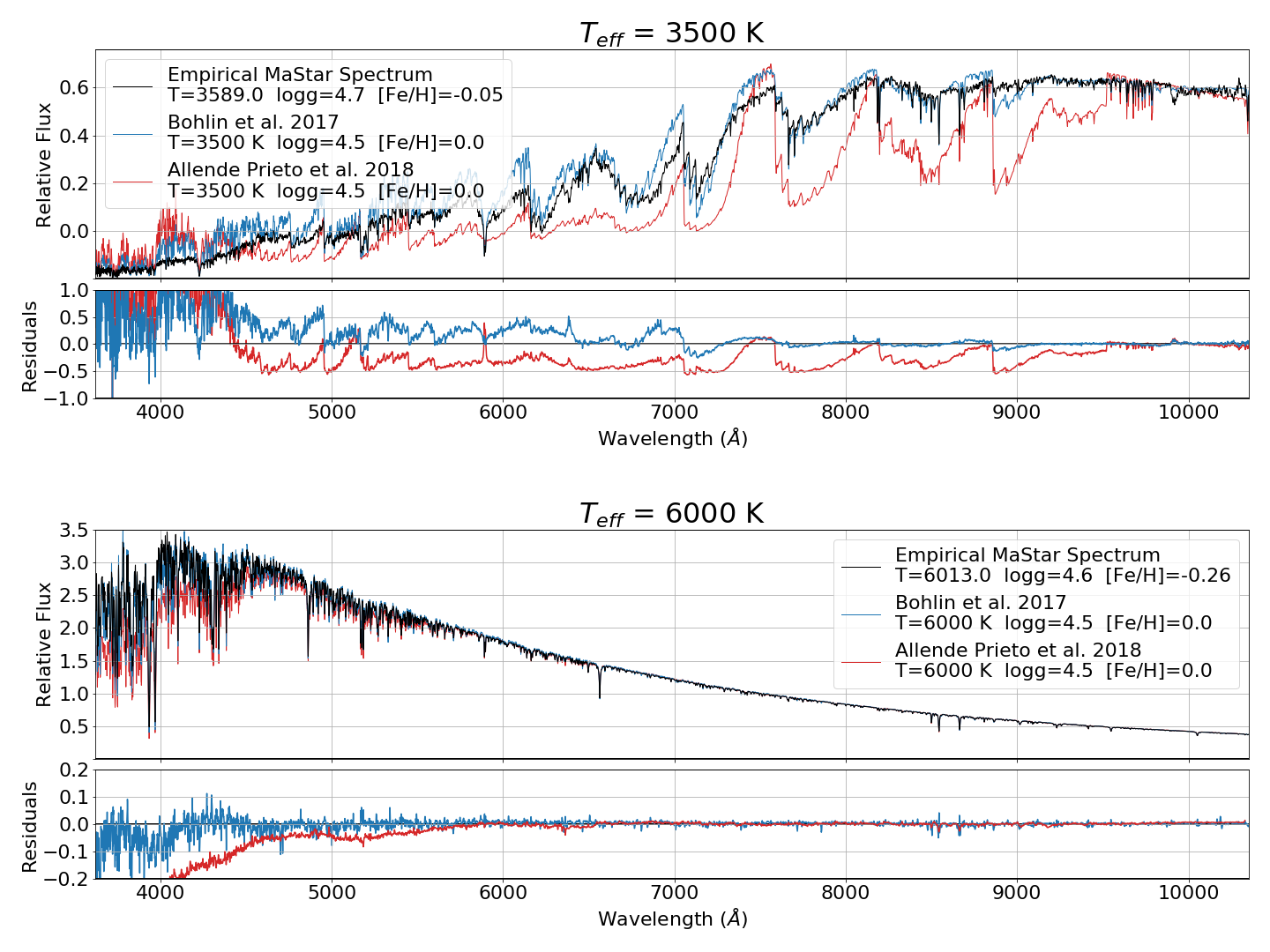}
    \caption{Stellar spectra comparing two different synthetic libraries, \cite{BOSZ} (blue) and \cite{Allende_Prieto_2018} (red), compared to an empirical MaStar spectrum (black) with similar parameters from ASPCAP. The top panel shows an M-dwarf star with stellar parameters $T_{eff}$ = 3500 K, $\log g$ = 4.5 dex, $[Fe/H]$ = $[\alpha/Fe]$ = 0.0. The bottom panel depicts an F-type star with $T_{eff}$ = 6000 K, $\log g$ = 4.5 dex, $[Fe/H]$ = $[\alpha/Fe]$ = 0.0. All Spectra have been normalized to a relative flux by dividing out a median value from a defined wavelength window. Residuals ($(f_{synth}-f_{MaStar})/f_{MaStar}$) are shown underneath each example.}
    \label{fig:synthcompare}
\end{figure}

To demonstrate this, two synthetic spectra of a M-dwarf star from different synthetic spectral libraries \citep{BOSZ, Allende_Prieto_2018} are shown in the top panel of Figure \ref{fig:synthcompare}, along with an observed M-dwarf spectrum from MaStar. The two synthetic spectra wildly differ from each other, despite representing the same set of parameters, most likely due to the more extensive molecular linelist used in \cite{Allende_Prieto_2018}. Neither spectrum is a close match to the observed star. For the hotter F-star spectra in the bottom panel, both synthetic libraries reasonably reproduce the observed spectrum. \cite{BOSZ} is a better match the the stellar continuum especially in bluer wavelengths ($\lambda \leq 5000$ \AA), because this library was used for the flux calibration in MaStar \citep{mastar2019}. \cite{Allende_Prieto_2018}, the library we adopt for our synthetic reference set, exhibits a lower amplitude of fluctuations in the residuals, indicating that disregarding continuum effects, it is a better match to the spectrum overall. The difference in continuum shape will not affect our results because we use continuum-normalized spectra during fitting. 

Of course, even the empirical half of the reference set relies heavily on synthetic spectra under its surface, since the ASPCAP parameters we adopt for the training set were derived themselves using a synthetic spectral grid. A key difference is that APOGEE spectra are much higher resolution than MaStar, cover infrared wavelengths, and significant effort was made to tune the line list for this spectral region.

By using a semi-empirical approach and adopting empirical spectra for our a large fraction of our reference set, we increase our ability to accurately fit the observed spectra for cool stars. Expanding our training set to include theoretical spectra in the warmer temperature regimes presents no disadvantage, as the model spectra quite closely match observations in this range, and allows us to fit the broader range of stellar temperatures relevant for
the full MaStar sample.  We expand on this discussion in Section \ref{section:Discussion:NN}.


\subsection{Procedure}

Our methodology is inspired by a combination of two algorithms called \textit{The Cannon} \citep{cannon1,cannon2} and \textit{The Payne} \citep{Ting_2019}. \textit{The Cannon} is a code that uses a data-driven approach to stellar parameter derivation, employing a quadratic model to predict the observed flux at a given pixel as a function of specified input parameters. \textit{The Payne} utilizes a similar framework to \textit{The Cannon}, but uses a theoretical spectral library to train a neural network as a generative flux model. Our routine uses observed spectra as a training set similar to the \textit{The Cannon}, and employs a neural network to model the flux as a function of the stellar parameters as in \textit{The Payne}. 

Using a neural network as a generative flux model presents several advantages over other methods for estimating the stellar parameters of a spectrum. First, many interpolation algorithms for determining stellar parameters require the training set to be a regularly-spaced 'grid' of spectra, which can quickly become computationally costly as the dimensions of the grid must be equal to the number of parameters one wishes to fit for. The training set for a generative model such as our neural network, \textit{The Payne}, and \textit{The Cannon}, can be both non-rectangular and smaller. As a result, these algorithms tend to be computationally fast and ideally suited for empirical training sets, which are inherently irregular.  Interpolation is not advised for empirical training sets, as any noise in the spectrum could be inadvertently captured in the interpolation. Generative flux models fit spectra as a continuous function of parameters which reduces the effect of such noise.

Additionally, neural networks are highly flexible and are able to capture variation in spectra across broad ranges of parameter space simultaneously \citep{Ting_2019}, unlike simpler polynomial models, which perform best within restricted ranges \citep{cannon1,cannon2,Rix2016}. The MaStar library was built to include as many types of stars as possible, so our parameter derivation pipeline must be flexible enough to simultaneously fit the spectrum of a 3000 K star as well as that of a 30,000 K star; a neural network is uniquely suited for this task.

We use the same architecture of neural network model as presented in \textit{The Payne}, using a “fully-connected" network with two hidden layers to predict the flux at each pixel as a function of the stellar parameters \citep[][Equation 1]{Ting_2019}. Practically, the neural network architecture is built through the Python \verb|PyTorch| module. Our reference set consists of both the empirical APOGEE-MaStar Overlap Sample and theoretical extension described in Sections \ref{section:Dataapogee} and \ref{section:DataSynth}.

This methodology can be broken down into a two-step process. The first step, called the \textit{training step}, trains a neural network model on a set of reference spectra for which the parameters are already known. The model can then be used to generate a spectrum for any given set of parameters. The training step relies on the underlying assumption that stellar flux can be modeled by a smooth, continuous function of the stellar parameters. Therefore, similar spectra have similar parameters, and similar parameters produce similar spectra.




Second is the \textit{application step}, where the model is applied to fit spectra with unknown parameters with a $\chi^2$-minimization procedure. The \textit{application step} is also sometimes referred to as the \textit{testing step} in other literature \citep[i.e.,][]{cannon1,cannon2,Ting_2019}. During the application step, individual spectra are fit through a $\chi^2$-minimization procedure. The starting guess for the search was calculated using Gaia photometry. A star's $(G_{BP}-G_{RP})$ color, corrected for interstellar extinction using a 3D dust map and Gaia-derived distance, and absolute $G$-band magnitude are used to provide a first estimate for effective temperature (\teff) and surface gravity (\logg) by identifying the closest point on a set of \verb|PARSEC| isochrones \citep{Bressan_2012}. If no Gaia photometry is available, which is only the case for $\approx1\%$ of MaStar targets, the search was conducted multiple times using different starting guesses, and the result with the lowest $\chi^2$ was adopted. The $\chi^2$-minimization is done in Python with the \verb|scipy.optimize.curve_fit| routine.

The search was restricted within the bounds [3000, 32,000] in temperature, [-1.0,5.5] in surface gravity, [-3.0,1.0] in metallicity, [-1.0,1.0] in alpha element abundance and [-1.1,1.1] in log($v_{micro}$). These bounds include slight extrapolation outside the parameter range covered by the training set. 

\subsection{Performance Assessment}
\label{section:Performance}

\begin{figure}
    \centering
    \includegraphics[width=\textwidth]{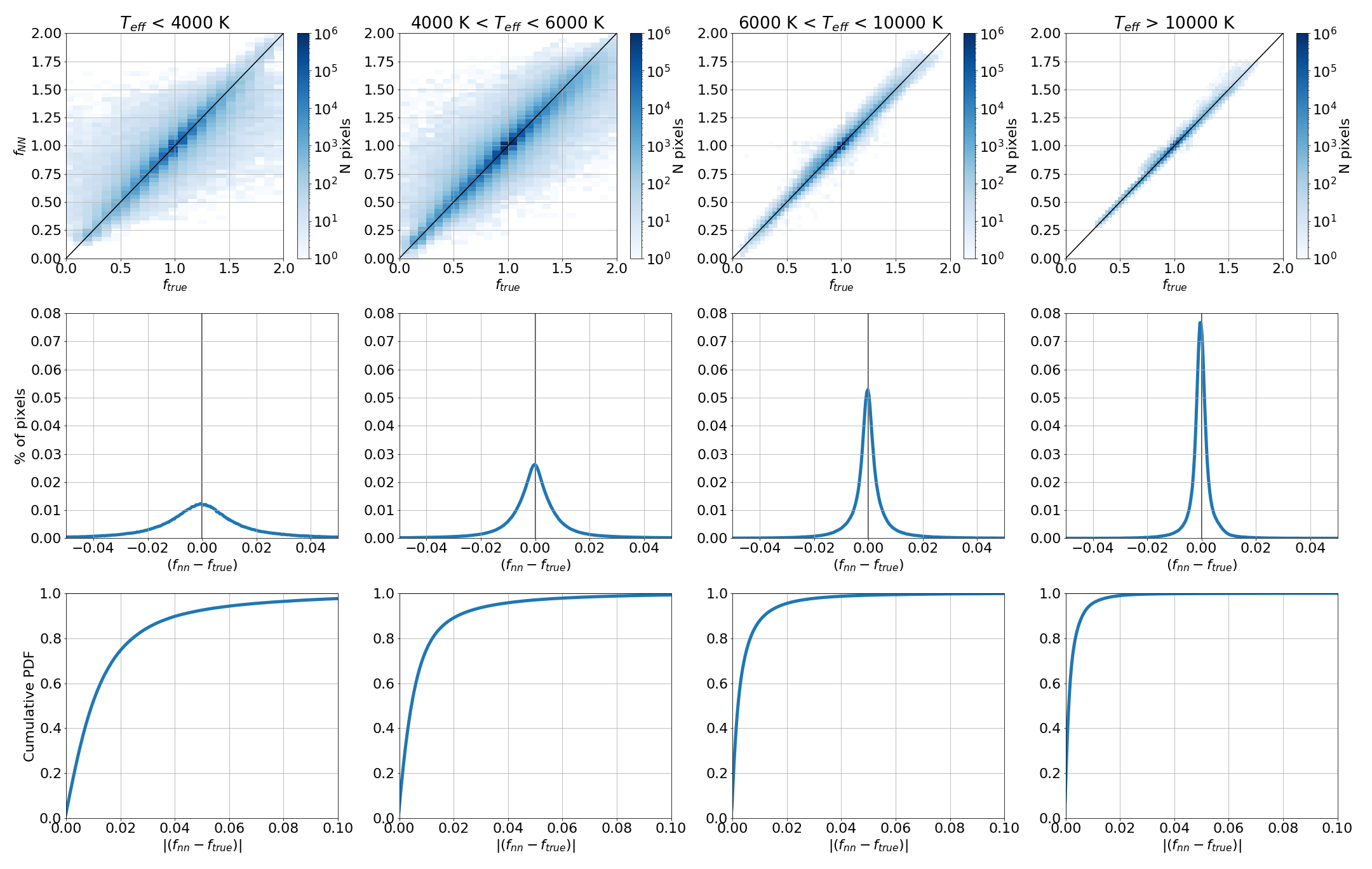}
    \caption{Performance of the neural network over a range of different temperature bins, from coolest (left column) to hottest (right column), demonstrating the neural network's ability to reproduce the flux of any given pixel in every spectrum of the training set. Top Row: one-to-one plots comparing the true flux values in the training set spectra ($f_{true}$) to the flux generated by the neural network ($f_{nn}$). Middle Row: Histograms showing the the distribution of spectral flux error $(f_{nn} - f_{true})$ calculated for every pixel in every spectrum in the training set. Bottom Row: the cumulative distribution function of the absolute spectral error $|(f_{nn} - f_{true})|$.}
    \label{fig:performance}
\end{figure}

The performance of the neural network as a generative model can be evaluated by testing how well the neural network can reproduce the spectra of the training set. This test is shown in Figure \ref{fig:performance}, plotted as a one-to-one comparison of the training set flux $f_{true}$ compared to the neural-network generated spectrum $f_{nn}$ (top row), the distribution of spectral errors (middle row), and the a cumulative distribution of spectral errors (bottom row) across different ranges of temperatures. At hotter temperatures, where the model was trained on noiseless synthetic spectra, the model performs extremely well with $99\%$ of all pixels reproduced within $2\%$ flux error as shown in the cumulative PDF. In cooler regimes, where the training set is dominated by empirical MaStar spectra, the presence of noise somewhat limits the usefulness of this metric in evaluating the neural network model, evidenced by the large scatter around the fiducial one-to-one line in the top row. Even so, the neural network is able to reproduce $90\%$ of pixels within $4\%$ of the observed flux at the coolest temperatures. If the performance at cool temperatures was at the same level as the hotter regimes, it would be a warning sign that the neural network was over-fitting the training set spectra. 


The previous exercise evaluated how well the neural network can reproduce the reference set spectra; a second experiment, a simple validation test, will examine the accuracy at which the neural network is capable of fitting parameters. During the training step of a typical neural network, a random $10\%$ of the reference set is reserved as a testing set to prevent over-fitting as the neural network converges on the best values to fit the training set. Here, we reserve an additional $10\%$ of the reference set for validation purposes in the application step here. After the neural network has been trained using the training set and testing set, the neural network is applied to fit for the parameters of the validation set. These parameters are compared to the 'true' values for the validation set spectra to assess the neural network's accuracy and precision.

\begin{figure}
    \centering
    \includegraphics[width=\textwidth]{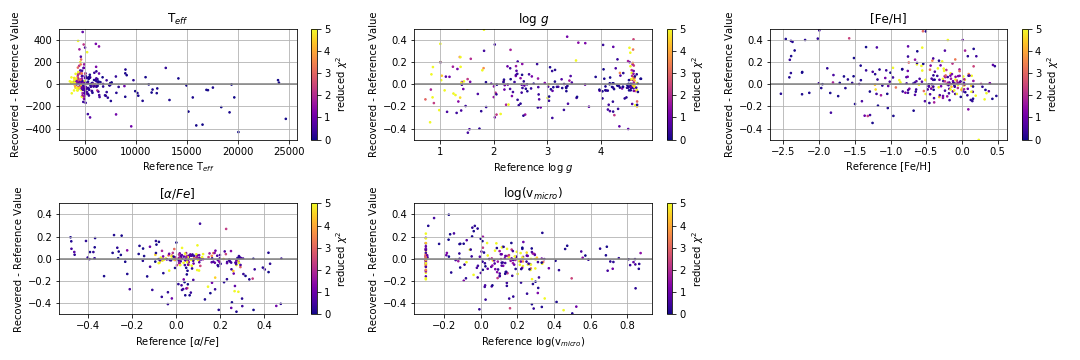}
    \caption{Results from a cross validation test, training the neural network on 90\% of the reference set and reserving the last 10\% for validation. Each panel shows a different parameter, highlighting the neural network's ability to accurately fit the remaining 10\% of spectra. Points are colored by the reduced $\chi^2$ value of the fits. Median offsets and scatter for each parameter are listed in Table \ref{tab:Xvalid}.}
    \label{fig:XValid}
\end{figure}

The results from the cross-validation are shown in Figure \ref{fig:XValid} and tabulated in Table \ref{tab:Xvalid}. The validation test offers insight into the systematic uncertainties that might be present in our final results, caused by variations in the stellar spectra that are not captured by the neural network. The recovered fit parameters are compared to the reference ASPCAP value, showing no significant offset bias in any parameter, and the scatter about the fiducial one-to-one line is reasonable within our expected precision; the scatter is comparable to the precision of the ASPCAP measurements and to the random uncertainties in our results described in Section \ref{section:uncertainties}.


\begin{table}[]
    \centering
    
    \begin{tabular}{| c | c | c | c | c |}
        \hline
        Parameter & Median Offset & Scatter & Median Offset & Scatter \\ 
        & (\teff $<$ 6000 K) &  (\teff $<$ 6000 K) & (\teff $>$ 6000 K) & (\teff $>$ 6000 K) \\
        \hline
        \hline
         $T_{eff}$ &  9.9 K & 169.5 K & 13.4 K & 739.9 K\\
         \hline
         $\log g$ &  0.0  & 0.29 & 0.02 & 0.17  \\
         \hline
         $[Fe/H]$ &  0.013  & 0.19 & 0.07 & 0.29 \\
         \hline
         $[\alpha / Fe]$ & 0.0  & 0.1 & 0.05 & 0.26  \\
         \hline
         $\log(v_{micro})$ &  0.02  & 0.18 & 0.02 & 0.29  \\
         \hline
    \end{tabular}
    \caption{Results from the validation test; showing the median offset and scatter (median absolute deviation) of the recovered results compared to the true values for five stellar parameters.}
    \label{tab:Xvalid}
\end{table}

\section{Results}
\label{section:Results}

\begin{figure}
    \centering
    \includegraphics[width=\textwidth]{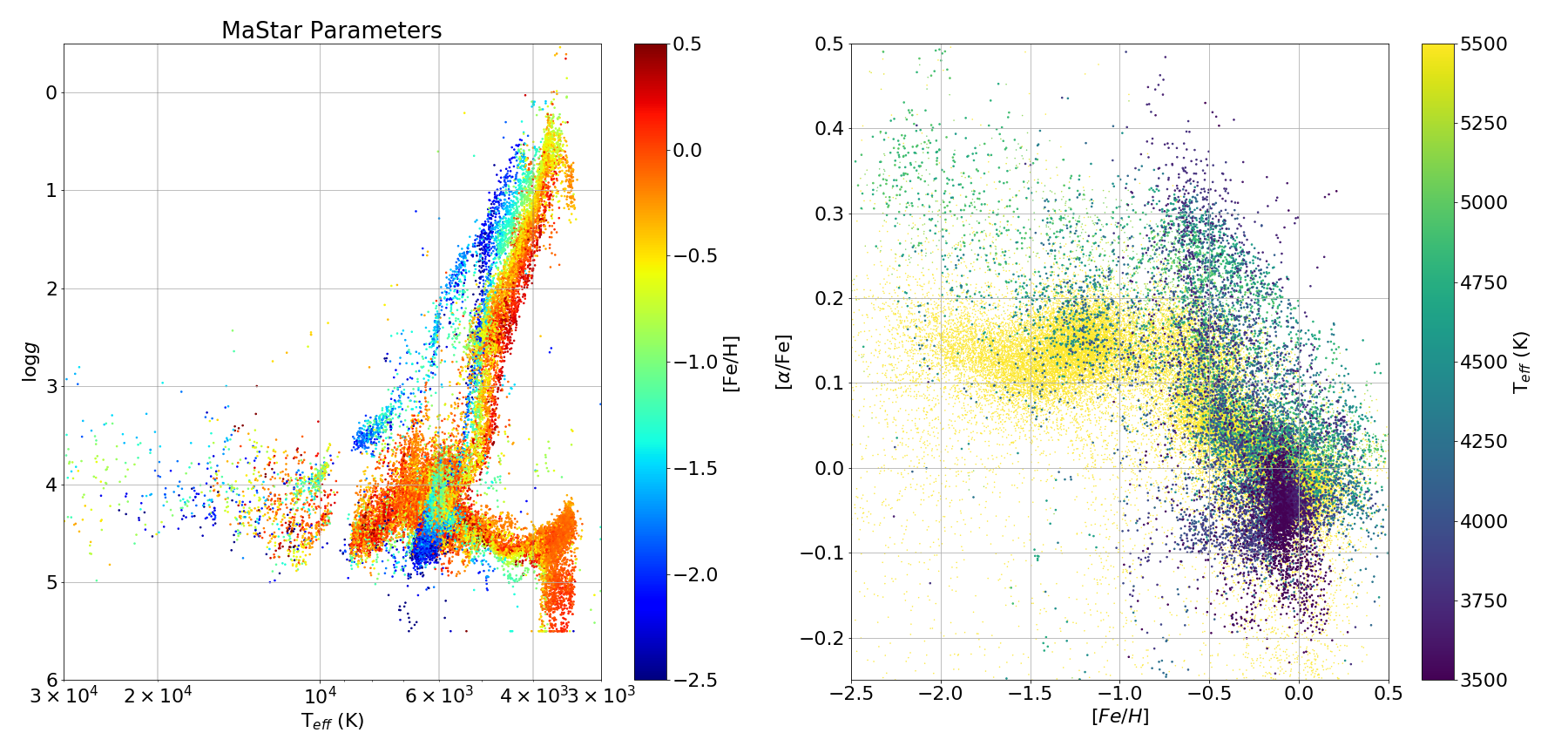}
    \caption{Parameters for the MaStar Stellar Library, presented in a Kiel diagram colored by $[Fe/H]$ (left) and the $[\alpha/Fe]-[Fe/H]$ plane colored by effective temperature (right).}
    \label{fig:FinalCMD}
\end{figure}

\subsection{Derived Parameters}

\begin{figure}
    \centering
    \includegraphics[width=0.5\textwidth]{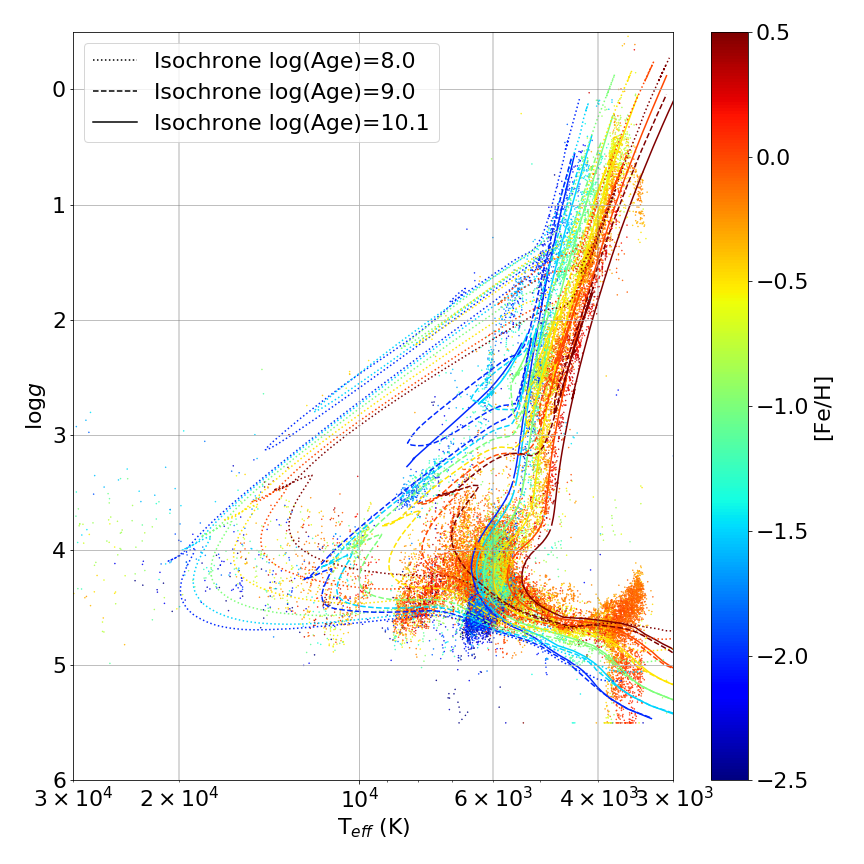}
    \caption{Parameters for the MaStar Stellar Library, plotted as a Kiel diagram colored by metallicity. Stellar isochrones from \cite{Bressan_2012} are overplotted for reference, with a variety of stellar ages (dotted, dashed, or straight lines) and different metalicities (-2.5 to 0.5 in steps of 0.5, with the same color scale).}
    \label{fig:CMD_isochr}
\end{figure}

The stellar parameters that we derived for the MaStar Stellar Library are shown in Figure \ref{fig:FinalCMD}. The neural-network model has derived a wide range of parameters for stars across many stellar evolutionary stages, reproducing much of the main sequence, red giant branch, and horizontal giant branch as expected from stellar evolution theory and isochrones \citep[][overplotted on the Kiel diagram in Figure \ref{fig:CMD_isochr}]{Bressan_2012}. The metal-poor clump of main sequence stars around $6000 < T_{eff} < 7500$ K are the standard stars which were targeted for flux calibration. The $[\alpha/Fe]-[Fe/H]$ in Figure \ref{fig:FinalCMD} plane shows the expected sequence of alpha-element abundances observed in the Milky Way \citep[e.g.,][]{Fuhrmann1998}. However, there are several strange features in our results that were likely artificially introduced by the neural network. These will be discussed in Section \ref{results:QC}.

\begin{figure}
    \centering
    \includegraphics[width=\textwidth]{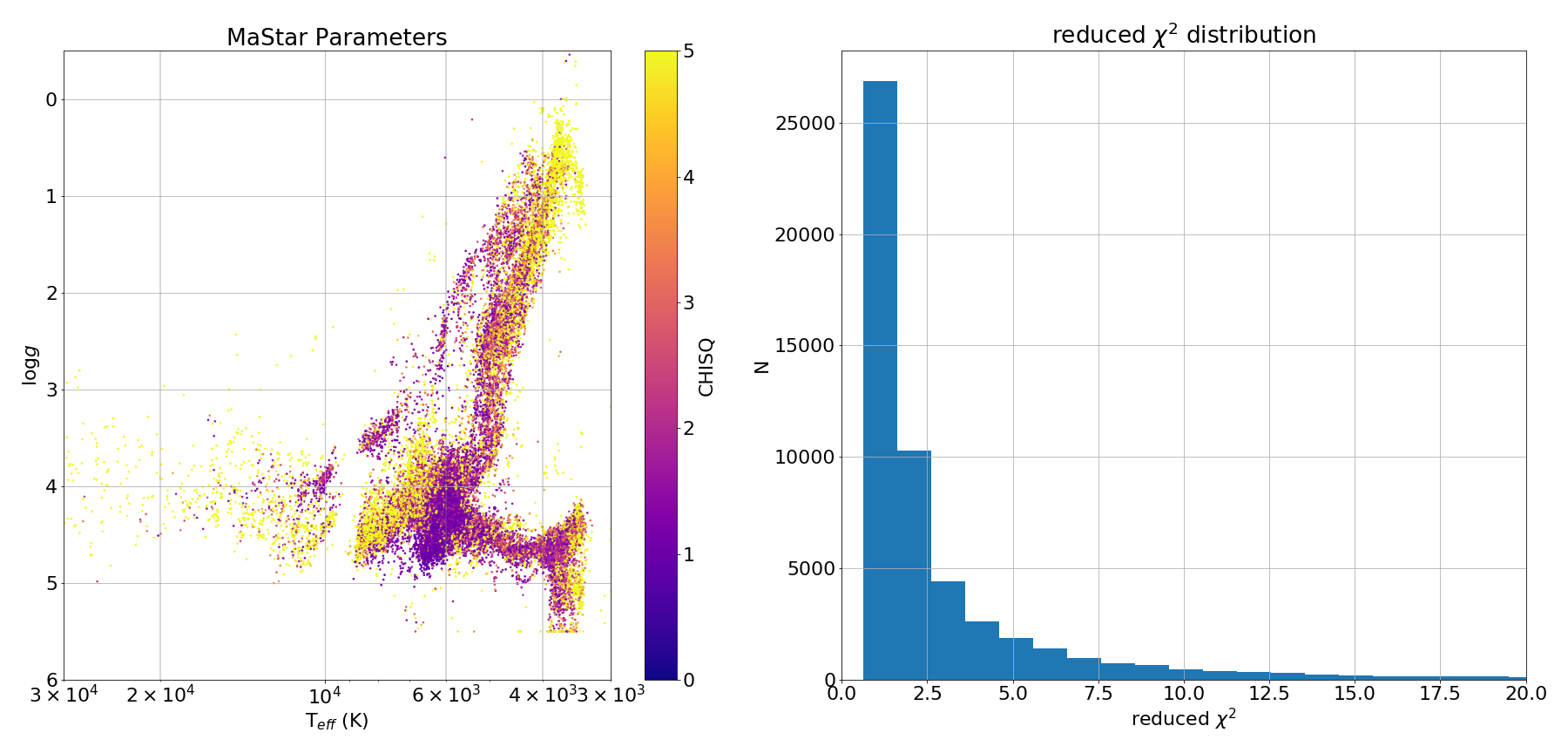}
    \caption{Left: Parameters for the MaStar Stellar library shown in a Kiel diagram, colored by the reduced $\chi^2$ value of our fits. Right: Histogram showing the distribution of reduced $\chi^2$ values.}
    \label{fig:CMDchi2}
\end{figure}

The reduced $\chi^2$ values for the fits are shown in a Kiel diagram and in a histogram in Figure \ref{fig:CMDchi2}. The Kiel diagram shows that the reduced $\chi^2$ varies with temperature; cool stars ($T_{eff} \leq 4000$ K) are expected to produce worse $\chi^2$ values given the large number of absorption features in the spectrum and high levels of variance in the training set spectra at cool temperatures. The reduced $\chi^2$ values are also higher for upper-main sequence stars, likely explained by the differences between the synthetic spectra in the training set and real observations as discussed in section \ref{section:MethodsMotivation}. The right panel shows that the majority fraction of our results are characterized by low $\chi^2$ values. Our fits produce reduced $\chi^{2} \leq 5 $ for over 80\% of the library. The median reduced $\chi^2$ value among all of our fits is 1.65.

\begin{figure}
    \centering
    \includegraphics[width=\textwidth]{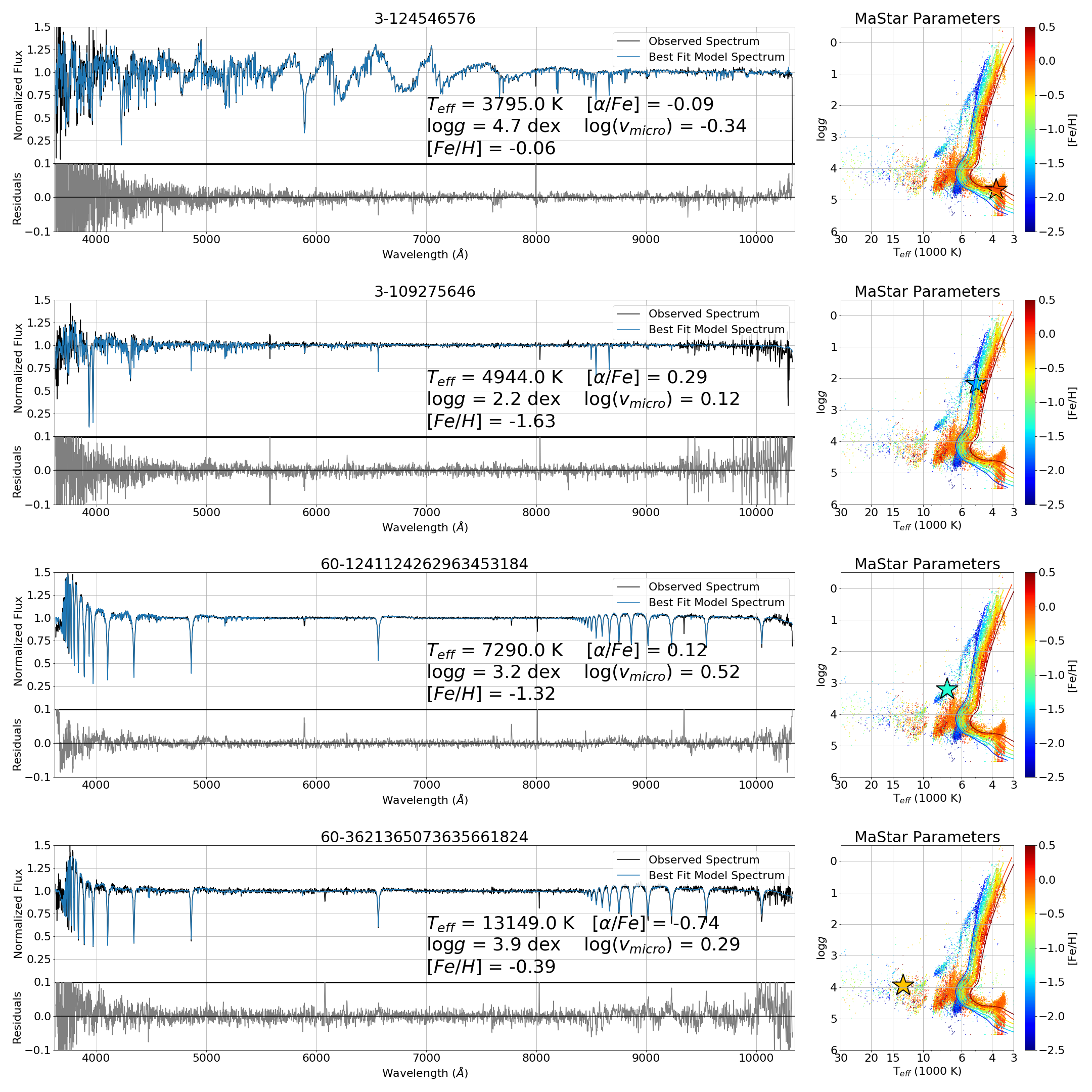}
    \caption{Example fits for four randomly-drawn spectra in different temperature bins from coolest (top row) to hottest (bottom row). For each star, its spectrum is shown on the left, with the observed spectrum ($f_{obs}$)in black and the best fit neural network model spectrum ($f_{nn}$) in blue. The residuals for the spectral fit ($f_{nn} - f_{obs}$) are shown in the panel underneath the spectrum. The derived parameters are shown in a Kiel diagram on the right.}
    \label{fig:FitExample}
\end{figure}

Four randomly-selected spectra within different ranges of parameters are shown in Figure \ref{fig:FitExample}, along with their best-fit model spectrum and residuals. Our method produces fits with low reduced $\chi^2$ values between the observed spectrum and neural network model spectrum for a broad range of temperature, surface gravities, and metallicities.  

\begin{figure}
    \centering
    \includegraphics[width=\textwidth]{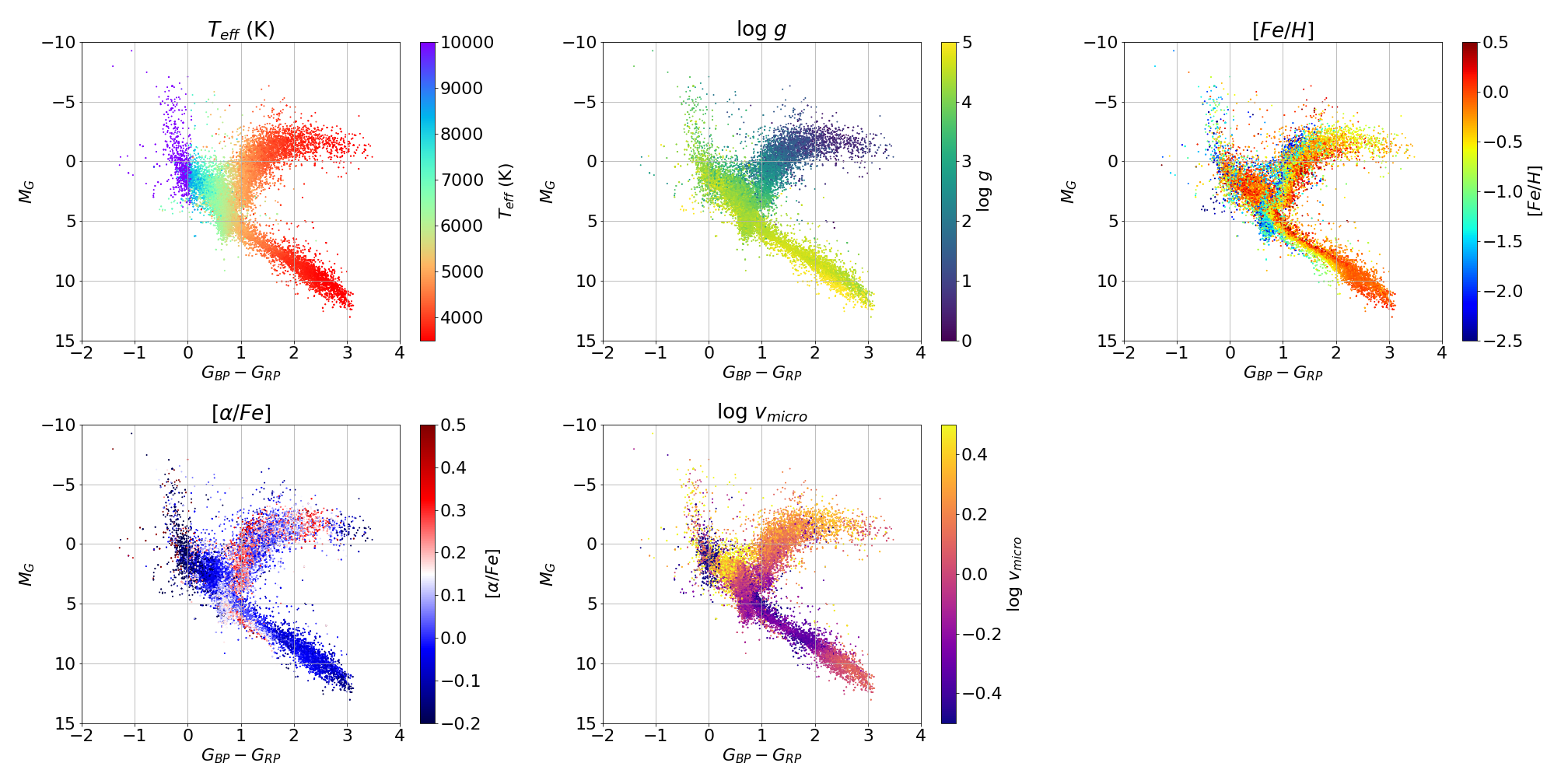}
    \caption{Color-Magnitude diagrams using Gaia $G$-band magnitude and dereddened ($G_{BP}-G_{RP}$) color. The points are colored by each of the parameters we derive in this work. Left to right and to top to bottom: $T_{eff}$, log $g$, $[Fe/H]$, $[\alpha/Fe]$, and log($v_{micro}$) }
    \label{fig:GaiaCMDs}
\end{figure}

In Figure \ref{fig:GaiaCMDs}, we plot our results against photometry from the Gaia survey by plotting a color-magnitude diagram using $M_G$ apparent magnitude and extinction-corrected ($G_{BP}-G_{RP}$) color using extinction values from a 3D dust map. Each panel is colored by a different parameter in our results. Temperature shows a smooth gradient with ($G_{BP}-G_{RP}$) color, with redder colors corresponding to cooler temperatures. In surface gravity, we find high values for dwarf stars that decrease gradually with $M_G$. For metallicity, a clear gradient can be seen in the red giant branch and the main sequence; metal-rich stars absorb more ultraviolet light through line-blanketing effects resulting in redder colors, and the presence of metals additionally contributes to internal structural changes in a star, resulting in cooler effective temperatures. In the Milky Way, alpha-element enhanced stars are mostly found in the metal-poor red giant branch, consistent with our results. Microturbulent velocity is found to be higher in giant stars, as expected from previous studies \citep[e.g.,][]{Montalban2007}.


\subsection{Quality Control}
\label{results:QC}

We provide parameters for \Nfits spectra, or 92\% of the full MaStar library. The remaining 8\% are omitted from our results due to quality concerns. Results were flagged as invalid and not included in our catalog if the the reduced $\chi^2$ of the fit was greater than 100. In a handful of other cases, the $\chi^2$-minimization procedure failed to converge on a solution. Further inspection showed that this usually happened due to some irregularity in the spectra (i.e., strong emission lines), or because the star was of a type that had no representation in our training set (i.e., white dwarfs or extreme HB stars). These were not included in our results as the fits did not converge. 

In some regions of parameter space, our results are likely affected by systematic artifacts introduced by the neural network from the fitting process and should be used with caution. For example, there is a void in in the main sequence around $T_{eff} = 10,000$ K that is likely artificial. This temperature range is where the hydrogen lines in a spectrum are at their peak strength. If the Hydrogen lines are deeper in the synthetic training set spectra than they are in the real stars, this would lead to a disparity causing the fits to preferentially fall on either side of this valley. Evidence of this effect will be shown in \cite{lazarz2021}. Our neural network is capable of accurately characterizing the training set spectra in this temperature range, shown in Section \ref{section:Performance}, but if the input spectra do not adequately match the observed spectra, voids like this could be introduced during the fits.


The lower-main sequence (\teff $\leq 3750$ K, \logg $\geq 4$) shows a wide distribution in $\log g$ that is not physical. This is likely caused by the distribution of stars in our training set. There are 22 dwarf stars ($0.6\%$) in the full reference set that fall in this range, meaning the neural network is receiving little information about the variation of spectra at cool temperatures. The extrapolation ability of a neural network is limited; it cannot be expected to perform well on stars that are not adequately represented in the training set. This same extrapolation issue probably caused the lack of metal-rich stars ($[Fe/H] \geq$  0.0) at the tip of the RGB in our results; there was not enough information in the training set at these low temperatures.

We also recommend caution regarding the metallicity and alpha-element abundance estimates for stars on the upper-main sequence (\teff $> 10,000$ K). We fit many of them as metal-poor, when they are expected to be younger stars that are more metal-rich. Metallicity is difficult to constrain in the spectrum of a hot star, since very few metal lines are present. This is recorded in the higher uncertainty values for these stars. Discrepancies between synthetic spectra and observed stars also contributes to the lower precision and accuracy of our results for the hot stars as mentioned in Section \ref{section:DataSynth}.

\subsection{External Comparison}
\label{section:Validation:Lit}

\begin{figure}
    \centering
    \includegraphics[width=\textwidth]{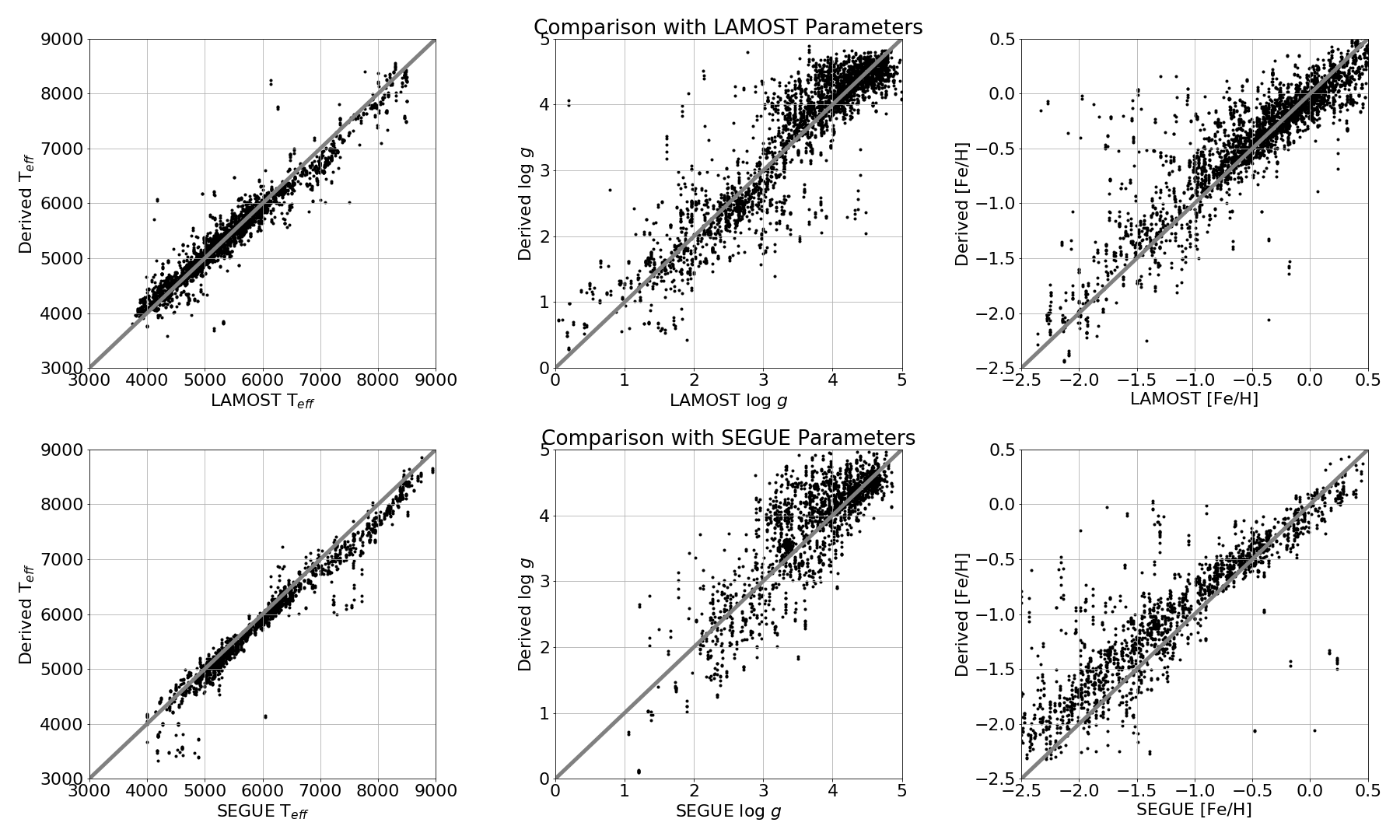}
    \caption{One-to-one comparison between our derived parameters and values from LAMOST (top row) and SEGUE (bottom row). Left column: \teff. Middle Column: \logg. Right column: \feh. The one-to-one line is drawn in gray as a guide.}
    \label{fig:LiteratureComparison}
\end{figure}

MaStar shares several thousand targets with other large stellar surveys, including LAMOST \citep[][$R \sim 1800; \lambda=3690-9100$ \AA]{LAMOST_Luo2015} and SEGUE \citep[][$R \sim 2000; \lambda=3800-9200$ \AA]{Lee_2008}. These catalogs provide independently-derived parameters for these stars, facilitating a direct comparison to help assess the quality of our results as shown in Figure \ref{fig:LiteratureComparison} and Table \ref{tab:LiteratureComparison}. Parameters from the LAMOST survey in \cite{LAMOST_Luo2015} were derived using two consecutive methods. An initial parameter estimate is obtained by interpolating a grid of ATLAS9 model spectra \citep{atlas9} to find the best match to the observed spectrum. This estimate is later used as a starting guess for the $\chi^2$-minimization procedure in the Universite de Lyon Spectroscopic analysis Software \citep[ULySS;][]{Koleva2009}, which interpolates spectra in the empirical ELODIE library \citep{Prugniel2001} to fit to the observed spectra. The SEGUE Stellar Parameter Pipeline \citep[SSPP;][]{Lee_2008} estimates stellar parameters through comparison to a theoretical spectral library \citep{AllendePrieto2006} calculated with \cite{Kurucz1993} ATLAS9 model atmospheres. The $\chi^2$ is minimized between the interpolated synthetic spectra and the observations. For high signal-to-noise spectra, they estimate precision to be 2\% in \teff, 0.2 dex in \logg, and 0.1 dex in \feh. These catalogs only provide effective temperature, surface gravity, and metallicity; comparison with our results in $[\alpha/Fe]$ and $\log v_{micro}$ is not possible here.

Our parameters are in good agreement with LAMOST, with offsets and scatter values reported in Table \ref{tab:LiteratureComparison}. In SEGUE, we find a larger systematic offset in temperature, with our results preferentially assigning cooler values than SEGUE. Our reported uncertainties are generally smaller than the scatter from this comparison. MaStar spectra usually have much higher S/N than LAMOST and SEGUE spectra due to the longer exposure time and more efficient instruments. Thus, we expect the large scatter in the comparison could have significant contributions from the stellar parameters uncertainties of LAMOST and SEGUE. 

\begin{table}[]
    \centering
    \begin{tabular}{| c | c | c | c |}
        \hline
         Parameter & Survey & Mean Offset & Scatter  \\
         \hline \hline
         $T_{eff}$ & LAMOST & 2.2 K & 304.9 K \\
         & SEGUE & 42 K & 538 K \\
         \hline
        $\log g$ & LAMOST & 0.04 dex &  0.44 dex \\
         & SEGUE & 0.11 dex & 0.50 dex \\
         \hline
        $[Fe/H]$ & LAMOST & 0.05 dex & 0.29 dex \\
         & SEGUE & 0.20 dex & 0.35 dex \\
         \hline
         
    \end{tabular}
    \caption{Comparison between our parameters and other literature catalogs; see also Figure \ref{fig:LiteratureComparison}. Showing the median offset and scatter (standard deviation) for three parameters in both catalogs.}
    \label{tab:LiteratureComparison}
\end{table}

\subsection{Precision of fits}

\label{section:uncertainties}

The precision of our parameters is reported in the uncertainties of the measurements. The uncertainties were calculated by deriving parameters for multiple observations of the same star independently, and calculating the unbiased estimator of standard deviation \citep{bailerjones_2017} for every pair of observations:

\begin{equation}
    \sigma_{X}^2 = \frac{(x_{1} - x_{2})^2}{(4/\pi)}
\end{equation}

where $x_1$ and $x_2$ are independent derivations of parameter $x$ from different observations of the same star. We found that the formal uncertainties derived from the $\chi^2$-minimization algorithm were generally smaller than the uncertainties calculated from multiple visits, so we use the multiple visit estimates to provide a more realistic measure of uncertainty. The distribution of $\log(\sigma_x)$ was modeled as a quadratic function of three parameters (\teff, \logg, \feh) and the median S/N ratio of a spectrum. This quadratic model was then used to calculate the final reported uncertainty values for each parameter for every spectrum. Typical uncertainty values are presented in Table \ref{tab:uncertainties}. 

These represent random uncertainties in our results, but not the systematic uncertainties. The validation test in Section \ref{section:Performance} provides an evaluation of the systematic uncertainties that are due to variations in the stellar spectra that are not captured by the neural network. In Section \ref{section:Discussion:LSF}, we will explore the systematic uncertainties caused by the variation in resolution among the MaStar Spectra.



\begin{table}[]
    \centering
    \begin{tabular}{|c|c|c|c|c|c|c|}
        \hline
         Temperature Bin & uncertainty estimate & $\sigma_{Teff}$ & $\sigma_{log g}$ & $\sigma_{[Fe/H]}$ & $\sigma_{[\alpha/Fe]}$ & $\sigma_{v_{micro}}$\\
         \hline
         \hline
         3000 K $<$ $T_{eff}$ $<$ 5000 K & Formal & 8.3 & 0.02 & 0.015 & 0.007 & 0.012 \\
         & Repeat Obs. & 8.2 & 0.02 & 0.014 & 0.006 & 0.012 \\
         \hline
         5000 K $<$ $T_{eff}$ $<$ 8000 K & Formal & 15.5 & 0.03 & 0.019 & 0.015 & 0.031 \\
         & Repeat Obs. & 16.2 & 0.03 & 0.020 & 0.0101 & 0.026 \\
         \hline
         8000 K $<$ $T_{eff}$ $<$ 12000 K & Formal & 30.7 & 0.02 & 0.050 & 0.035 & 0.049 \\
         & Repeat Obs. & 28.7 & 0.04 & 0.030 & 0.011 & 0.038 \\
        \hline
         12000 K $<$ $T_{eff}$ $<$ 20000 K & Formal & 96.1 & 0.03 & 0.104 & 0.067 & 0.106 \\
         & Repeat Obs. & 42.9 & 0.05 & 0.089 & 0.028 & 0.068 \\
         \hline
         20000 K $<$ $T_{eff}$ $<$ 35000 K & Formal & 478.74 & 0.04 & 0.15 & 0.077 & 0.114 \\
         & Repeat Obs. & 101.0 & 0.15 & 0.085 & 0.087 & 0.170 \\
         \hline
         
    \end{tabular}
    \caption{Median precision for each parameter in different temperature bins. “Formal" uncertanties were calculated from the covariance matrix of $\chi^2$-minimzation procedure. “Repeat Obs" errors were calculated from the standard error between multiple observations of the same stars, and fit with a quadratic function.}
    \label{tab:uncertainties}
\end{table}




\section{Discussion}
\label{section:Discussion}

\subsection{Effects of Varying Resolution}
\label{section:Discussion:LSF}

As mentioned in Section \ref{section:DataSpectra}, the resolution of MaStar observations varies across wavelength and across optical fibers. Every spectrum in MaStar's main data product has a unique resolution vector. In the synthetic part of our training set (described in section \ref{section:DataSynth}), we use the median resolution as a function of wavelength to match our training set to the data. Beyond that, we do nothing to account for this variation in resolution in our parameter results, as it would be too computationally expensive to convolve a unique training set to match the unique resolution of every MaStar spectrum, which would be needed to properly account for this effect. 

The uncertainty estimates calculated from multiple observations of the same star in section \ref{section:uncertainties} should account for the effects of resolution variation, since the multiple observations have different resolution as a function of wavelength. Here, we perform a more controlled test to examine the effects that the varying resolution has on our parameter results. 

\begin{figure}
    \centering
    \includegraphics[width=\textwidth]{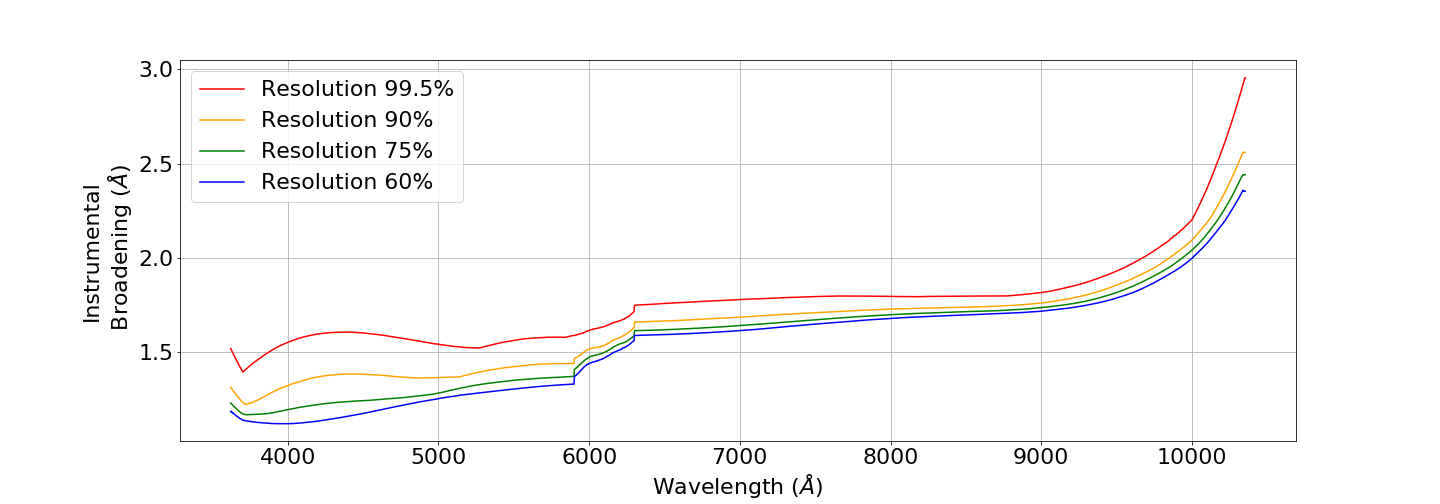}
    \caption{The resolution as a function of wavelength for each of the resolution-homogenzied MaStar catalogs released; the 99.5\% resolution catalog (red), the 90\% resolution catalog (yellow), the 75\% resolution catalog (green), and the 60\% resolution catalog (blue). The instrumental broadening value is a Gaussian $\sigma$ in units of Angstroms.}
    \label{fig:lsfs_homog}
\end{figure}

As a side product of the survey, MaStar has released several catalogs of the library that have been resolution homogenized as described in \cite{mastar2021}. Four different resolution vectors were selected, matching different percentiles of the resolution distribution of the full library; 60\%, 75\%, 90\%, and 99.5\%. Spectra are not de-convolved to higher resolution, so each homogenized catalog represents only a fraction of the library, which have been convolved to match the target resolution. For example, the 60-percentile resolution curve is defined by the 60-percentile in the resolution distribution among all good visit spectra at each wavelength, counting from the sharpest to the coarsest. The 60-percentile catalog contains about 14\% of the good visits spectra that have better resolution than this curve at all wavelengths between 3700 \AA and 10,000 \AA. In this file, all spectra have been convolved to match the 60-percentile resolution curve. The resolution vector for each of these catalogs is shown in Figure \ref{fig:lsfs_homog}.

\begin{figure}
    \centering
    \includegraphics[width=0.85\textwidth]{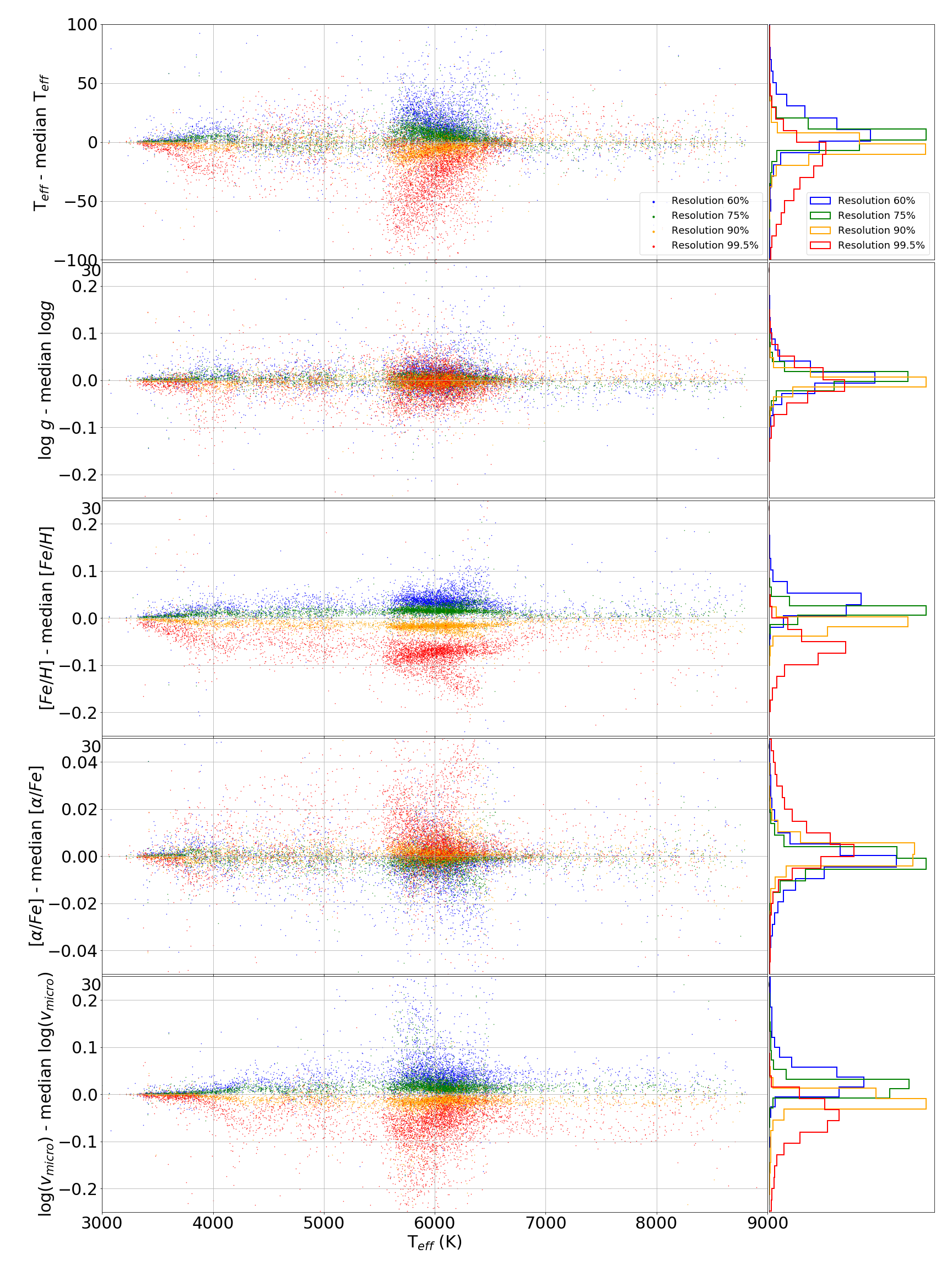}
    \caption{Examining the spread in parameters results while varying the resolution of spectra fit with our method. Each row represents a different parameter, from top to bottom: \teff, \logg, \feh, \alphafe, \vmicro. Each color represents the results from one of the four resolution catalogs: The 99.5\% resolution catalog (red), the 90\% resolution catalog (yellow), the 75\% resolution catalog (green), and the 60\% resolution catalog (blue). The median is calculated among the results from the four catalogs. The right column is the same thing flattened into a histogram.}
    \label{fig:lsf}
\end{figure}

We run each resolution percentile catalog through the pipeline independently. As a result, we calculate four sets of parameters for every star, derived from that star's spectrum in each of the four parameter catalogs. In Figure \ref{fig:lsf}, we explore the effects that the varying resolution has on the parameter results. Each star has four measurements of $T_{eff}$; each one calculated using its spectrum from the different resolution catalogs. The median among these four $T_{eff}$ measurements is used as a baseline to quantify the deviation each catalog presents compared to each other; this spread in derived $T_{eff}$ is an estimate of the systematic uncertainty in $T_{eff}$ that might be present in our results due to not accounting for varying resolution vectors. The same is applied to the other parameters.

The resolution has a small (but non-negligible) effect on the derivation of effective temperature, surface gravity, and alpha-element abundances. The parameters from the highest resolution ($60\%$) and lowest resolution ($99.5\%$), catalogs show more spread than the mid-resolution catalogs, most significant for temperatures around $T_{eff} \approx 6000 K$. In metallicity and microturbulence, resolution seems to have a greater effect on the derived parameters. The highest resolution catalog ($60\%$) preferentially fits to higher metallicities and microturbulence. Qualitatively, varying the resolution will change the shape of the absorption features, and absorption features are most sensitive to those parameters. A lower-resolution spectrum will have shallower (but broader) absorption lines, therefore masquerading as a lower-metallicity spectrum. In other words, when the fitting routine has no information on resolution, it attempts to explain the differences in line shape through differences in metallicity or microturbulence instead. From this, we conclude that our main pipeline may be under-estimating the metallicity and microturbulence for stars in the original catalog with lower-than-average resolution. Still, this effect is quite small, offsetting the metallicity derivation by approximately $0.1$ dex in the worst of cases. 

\subsection{Semi-empirical vs. Synthetic Neural Network model}
\label{section:Discussion:NN}

Data-driven approaches to determining stellar parameters, like this work, \textit{The Cannon}, and \cite{Chen_2020} use observations of real stars with known parameters as the reference set for estimating parameters. This is advantageous compared to the use of a synthetic library, as discussed in section \ref{section:MethodsMotivation}, as certain types of stars tend to be inadequately modeled. Using a subset of the full data to be fitted has the further advantage of less preparation work: the resolution, wavelength range, and noise levels in the reference set are already identical to the full data set. 

Here, we further justify our choice of using a data-driven model by comparing the results from two neural network models: the "semi-empirical" model trained on the empirical spectra plus the theoretical extension, which produced the results in Section \ref{section:Results}, and a second "synthetic model" trained only on synthetic spectra from \cite{Allende_Prieto_2018}. The training set for the synthetic model has an identical parameter distribution to the semi-empirical model. The synthetic spectral grid from \cite{Allende_Prieto_2018} was interpolated using the code \verb|FERRE| in pure interpolation mode \citep{ferre} to generate the needed synthetic spectra to match the distribution of empirical spectra. The architecture of both neural networks is also identical. The only difference between the two models is the training set spectra. 

\begin{figure}
    \centering
    \includegraphics[width=\textwidth]{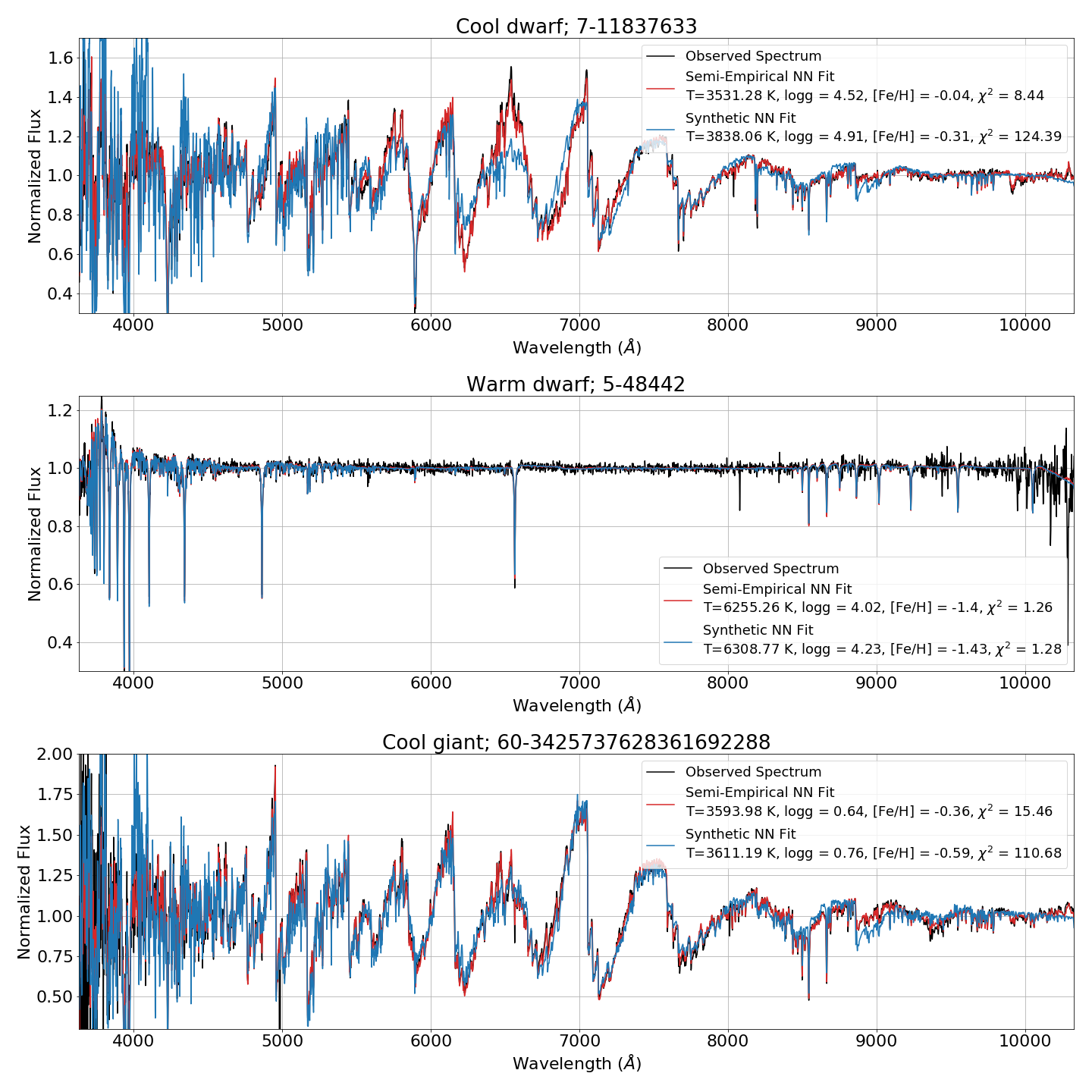}
    \caption{Comparison between fits for an observed MaStar spectrum (black line) with results from the data-driven (semi-empirical) neural network model (red) and a model-spectra-driven (synthetic) neural network model (blue) described in Section \ref{section:Discussion:NN}. Results are shown for three different stars in different parameter regimes: a cool dwarf (top), a warm dwarf (middle) and a cool giant (bottom). Select derived parameters for both models are listed in the legend. The semi-empirical model is a better fit to the observed spectrum in all three cases, with the largest discrepancy seen in the cool dwarf star. This motivates our choice to use the semi-empirical model for our final results presented in Section \ref{section:Results}.}
    \label{fig:empvsynth_specs}
\end{figure}

Example fits for three MaStar spectra are shown in Figure \ref{fig:empvsynth_specs} in different regimes of parameter space; a cool dwarf ($T_{eff} \approx 3500$ K, log$ g \approx$ 4.5), a warm dwarf ($T_{eff} \approx 6000$ K, log$ g \approx$ 4.0), and a cool giant star ($T_{eff} \approx 3500$ K, log$ g \approx$ 0.5). The best-fit spectrum from the semi-empirical neural netork model is shown in red, and the best-fit spectrum for the synthetic neural network model in blue. For all three stars, the observed spectrum (black line) is better fit by the semi-empirical model, with a lower reduced $\chi^2$ than the best fit from the synthetic model. The biggest discrepancy between the two models is obvious in the cool dwarf spectrum; the synthetic neural network does not well reproduce the peaks in the spectrum from the TiO bands. The derived parameters from the synthetic model are also significantly different than the parameters derived from the semi-empirical model; a difference in temperature of $\Delta T_{eff}$ $>$ 300 K, in logg of $>$ 0.4 dex, and in $[Fe/H]$ of almost 0.3 dex. For the warm dwarf example star, there is little difference between the results from the two neural network models. 

\begin{figure}[H]
    \centering
    \includegraphics[width=\textwidth]{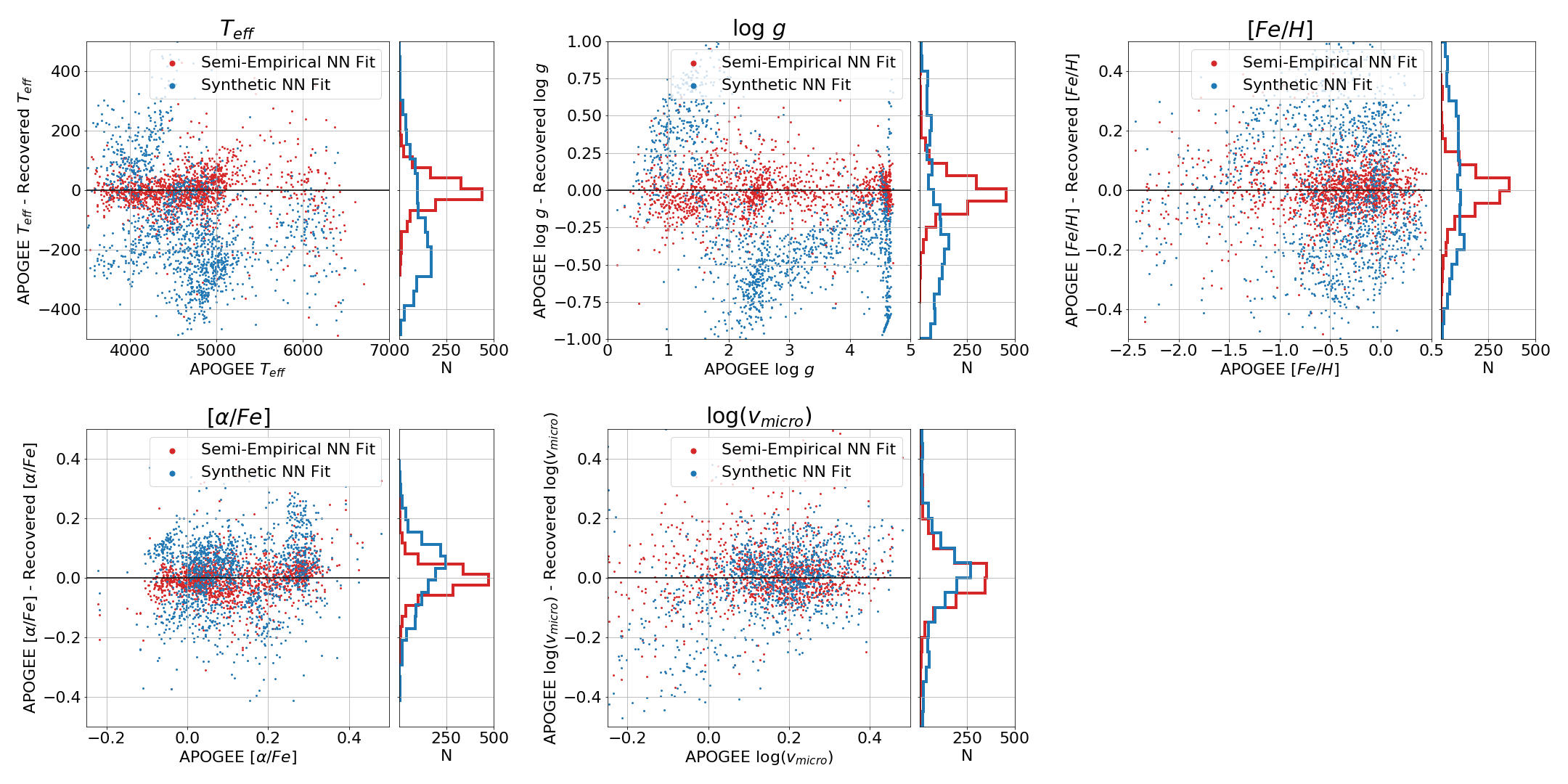}
    \caption{Parameter results from the semi-empirical neural network model (red points) and the synthetically-trained neural network model (blue points) discussed in Section \ref{section:Discussion:NN}, both compared to the ASPCAP parameters for the training set stars, shown as a scatter plot (left panels) and in a histogram (right panels). The synthetic neural network has larger scatter in all parameters, most consequential for cool dwarf stars where the biggest difference between synthetic and empirical spectra is observed. This motivates our choice to use the semi-empirical model for our final results presented in Section \ref{section:Results}.}
    \label{fig:empvsynth_params}
\end{figure}

The parameters derived by the two models differ significantly, as shown in Figure \ref{fig:empvsynth_params} in a validation test. The results from both models are compared against the `true' values, the ASPCAP parameters used to train. While the semi-empirical model produces results generally close to the ASPCAP parameters, the synthetic model has large scatter, particularly for cool stars ($T_{eff} < 5000$ K) where the temperature fits can be off by hundreds of degrees, and the surface gravity predictions off by up to 1.0 dex. 

The conclusion of this test should not be to avoid synthetic spectra entirely during parameter derivation. As mentioned in Section \ref{section:MethodsMotivation}, even the empirical half of our reference set is not truly independent of synthetic spectra because models were used to derive the parameters in ASPCAP. This test demonstrates that in this situation, at lower resolution and in optical wavelengths, the semi-empirical training set produces better parameter results. 

\section{Conclusions}
\label{section:Conclusions}

The MaNGA Stellar Library (MaStar) is the largest collection of uniform empirical stellar spectra published to date. The wide variety of stars contained in the catalog will be useful for many astronomical applications, from estimating the parameters of individual stars to the stellar population modeling of galaxies \citep{maraston2020}. The spectra in the MaStar Stellar library represent many different stars across evolutionary states and with a wide range of stellar properties as shown in Figure \ref{fig:FinalCMD}:

\begin{itemize}
    \item $3,000$ K $\lessapprox T_{eff} \lessapprox 32,000$ K
    \item $-0.75$ dex $\lessapprox \log g \lessapprox 5.5$ dex
    \item $-3.0$ dex $\lessapprox [Fe/H] \lessapprox 0.5$ dex
    \item $-0.25$ dex $\lessapprox [\alpha/Fe] \lessapprox 0.75$ dex
    \item $-1.0$ $\lessapprox \log (v_{micro}) \lessapprox 1.0$ log(km/s)
\end{itemize}

In this paper, we derive five stellar parameters: effective temperature (\teff), surface gravity (\logg), metallicity (\feh), alpha-element abundance (\alphafe), and microturbulent velocity (\vmicro) for \Nfits spectra in the MaStar library using a machine learning approach. For the reference set, we select a subset of \Ntraining MaStar spectra with precise parameters from ASPCAP. The training set is extended to cover hotter temperature regimes ($T_{eff} >$ 6000 K) with an additional \Ntraining randomly-selected synthetic spectra. We performed a number of quality tests to show that the neural network model trained on this reference set performs well, reproducing over 95\% of pixels within 2\% flux accuracy. 


The semi-empirical neural network model has derived stellar parameters for a wide range of stars across evolutionary states. The resulting fits have typical reduced $\chi^2$ values below 5.0. The precision of our results was evaluated by comparing the results across multiple observations of the same star, and our reported uncertainties updated to match that distribution. We report typical precision within 15 K in \teff,  0.03 dex in \logg, 0.02 dex in \feh, 0.2 dex in \alphafe, and 0.02 dex in \vmicro. Our parameters are in good agreement with other published catalogs including LAMOST and SEGUE.

A total of four efforts in the MaStar collaboration, including the work presented here, have derived parameters for the final release of the MaStar library utilizing different methodologies. The final catalog of all four sets of stellar parameters, a comparison between them, and a combined parameter set will be presented in \cite{mastar2021}. The individual procedure for each method not presented here will be detailed in \cite{Chen2021}, \cite{hill2021}, and \cite{lazarz2021}. As the four independent methods each have unique strengths and weaknesses, the combination provides a high-quality set of stellar parameters for the MaStar stellar library. 

The MaStar parameter catalog containing our results will be available through the official release of the SDSS-IV survey as a value-added catalog, for download through the SDSS-IV Science Archive Server$^1$.

\section*{Acknowledgements}

JI, JH, and DB acknowledge support from NSF grant AST-1715670.

Funding for the Sloan Digital Sky Survey IV has been provided by the Alfred P. Sloan Foundation, the U.S. Department of Energy Office of Science, and the Participating Institutions. SDSS acknowledges support and resources from the Center for High-Performance Computing at the University of Utah. The SDSS web site is www.sdss.org.

SDSS is managed by the Astrophysical Research Consortium for the Participating Institutions of the SDSS Collaboration including the Brazilian Participation Group, the Carnegie Institution for Science, Carnegie Mellon University, Center for Astrophysics | Harvard \& Smithsonian (CfA), the Chilean Participation Group, the French Participation Group, Instituto de Astrofísica de Canarias, The Johns Hopkins University, Kavli Institute for the Physics and Mathematics of the Universe (IPMU) / University of Tokyo, the Korean Participation Group, Lawrence Berkeley National Laboratory, Leibniz Institut für Astrophysik Potsdam (AIP), Max-Planck-Institut für Astronomie (MPIA Heidelberg), Max-Planck-Institut für Astrophysik (MPA Garching), Max-Planck-Institut für Extraterrestrische Physik (MPE), National Astronomical Observatories of China, New Mexico State University, New York University, University of Notre Dame, Observatório Nacional / MCTI, The Ohio State University, Pennsylvania State University, Shanghai Astronomical Observatory, United Kingdom Participation Group, Universidad Nacional Autónoma de México, University of Arizona, University of Colorado Boulder, University of Oxford, University of Portsmouth, University of Utah, University of Virginia, University of Washington, University of Wisconsin, Vanderbilt University, and Yale University.
\newline
\newline
$^1$\url{https://data.sdss.org/sas/dr17/manga/spectro/mastar/v3_1_1/v1_7_7/vac/parameters/v1/mastar-goodvisits-v3_1_1-v1_7_7-params-v1.fits}

\bibliography{biblio}{}

\begin{thebibliography}{}
\expandafter\ifx\csname natexlab\endcsname\relax\def\natexlab#1{#1}\fi
\providecommand{\url}[1]{\href{#1}{#1}}
\providecommand{\dodoi}[1]{doi:~\href{http://doi.org/#1}{\nolinkurl{#1}}}
\providecommand{\doeprint}[1]{\href{http://ascl.net/#1}{\nolinkurl{http://ascl.net/#1}}}
\providecommand{\doarXiv}[1]{\href{https://arxiv.org/abs/#1}{\nolinkurl{https://arxiv.org/abs/#1}}}

\bibitem[{{Allende-Prieto} \& {Apogee Team}(2015)}]{ferre}
{Allende-Prieto}, C., \& {Apogee Team}. 2015, in American Astronomical Society
  Meeting Abstracts, Vol. 225, American Astronomical Society Meeting Abstracts
  \#225, 422.07

\bibitem[{{Allende Prieto} {et~al.}(2006){Allende Prieto}, {Beers}, {Wilhelm},
  {Newberg}, {Rockosi}, {Yanny}, \& {Lee}}]{AllendePrieto2006}
{Allende Prieto}, C., {Beers}, T.~C., {Wilhelm}, R., {et~al.} 2006, \apj, 636,
  804, \dodoi{10.1086/498131}

\bibitem[{Allende~Prieto {et~al.}(2018)Allende~Prieto, Koesterke, Hubeny,
  Bautista, Barklem, \& Nahar}]{Allende_Prieto_2018}
Allende~Prieto, C., Koesterke, L., Hubeny, I., {et~al.} 2018, Astronomy \&
  Astrophysics, 618, A25, \dodoi{10.1051/0004-6361/201732484}

\bibitem[{Bailer-Jones(2017)}]{bailerjones_2017}
Bailer-Jones, C. A.~L. 2017, Practical Bayesian Inference: A Primer for
  Physical Scientists (Cambridge University Press),
  \dodoi{10.1017/9781108123891}

\bibitem[{{Berger} {et~al.}(2020){Berger}, {Huber}, {van Saders}, {Gaidos},
  {Tayar}, \& {Kraus}}]{Berger2020}
{Berger}, T.~A., {Huber}, D., {van Saders}, J.~L., {et~al.} 2020, \aj, 159,
  280, \dodoi{10.3847/1538-3881/159/6/280}

\bibitem[{{Blanton} {et~al.}(2017){Blanton}, {Bershady}, {Abolfathi},
  {Albareti}, {Allende Prieto}, {Almeida}, {Alonso-Garc{\'\i}a}, {Anders},
  {Anderson}, {Andrews}, {Aquino-Ort{\'\i}z}, {Arag{\'o}n-Salamanca},
  {Argudo-Fern{\'a}ndez}, {Armengaud}, {Aubourg}, {Avila-Reese}, {Badenes},
  {Bailey}, {Barger}, {Barrera-Ballesteros}, {Bartosz}, {Bates}, {Baumgarten},
  {Bautista}, {Beaton}, {Beers}, {Belfiore}, {Bender}, {Berlind}, {Bernardi},
  {Beutler}, {Bird}, {Bizyaev}, {Blanc}, {Blomqvist}, {Bolton}, {Boquien},
  {Borissova}, {van den Bosch}, {Bovy}, {Brandt}, {Brinkmann}, {Brownstein},
  {Bundy}, {Burgasser}, {Burtin}, {Busca}, {Cappellari}, {Delgado Carigi},
  {Carlberg}, {Carnero Rosell}, {Carrera}, {Chanover}, {Cherinka}, {Cheung},
  {G{\'o}mez Maqueo Chew}, {Chiappini}, {Choi}, {Chojnowski}, {Chuang},
  {Chung}, {Cirolini}, {Clerc}, {Cohen}, {Comparat}, {da Costa}, {Cousinou},
  {Covey}, {Crane}, {Croft}, {Cruz-Gonzalez}, {Garrido Cuadra}, {Cunha},
  {Damke}, {Darling}, {Davies}, {Dawson}, {de la Macorra}, {Dell'Agli}, {De
  Lee}, {Delubac}, {Di Mille}, {Diamond-Stanic}, {Cano-D{\'\i}az}, {Donor},
  {Downes}, {Drory}, {du Mas des Bourboux}, {Duckworth}, {Dwelly}, {Dyer},
  {Ebelke}, {Eigenbrot}, {Eisenstein}, {Emsellem}, {Eracleous}, {Escoffier},
  {Evans}, {Fan}, {Fern{\'a}ndez-Alvar}, {Fernandez-Trincado}, {Feuillet},
  {Finoguenov}, {Fleming}, {Font-Ribera}, {Fredrickson}, {Freischlad},
  {Frinchaboy}, {Fuentes}, {Galbany}, {Garcia-Dias},
  {Garc{\'\i}a-Hern{\'a}ndez}, {Gaulme}, {Geisler}, {Gelfand},
  {Gil-Mar{\'\i}n}, {Gillespie}, {Goddard}, {Gonzalez-Perez}, {Grabowski},
  {Green}, {Grier}, {Gunn}, {Guo}, {Guy}, {Hagen}, {Hahn}, {Hall}, {Harding},
  {Hasselquist}, {Hawley}, {Hearty}, {Gonzalez Hern{\'a}ndez}, {Ho}, {Hogg},
  {Holley-Bockelmann}, {Holtzman}, {Holzer}, {Huehnerhoff}, {Hutchinson},
  {Hwang}, {Ibarra-Medel}, {da Silva Ilha}, {Ivans}, {Ivory}, {Jackson},
  {Jensen}, {Johnson}, {Jones}, {J{\"o}nsson}, {Jullo}, {Kamble}, {Kinemuchi},
  {Kirkby}, {Kitaura}, {Klaene}, {Knapp}, {Kneib}, {Kollmeier}, {Lacerna},
  {Lane}, {Lang}, {Law}, {Lazarz}, {Lee}, {Le Goff}, {Liang}, {Li}, {Li},
  {Lian}, {Lima}, {Lin}, {Lin}, {Bertran de Lis}, {Liu}, {de Icaza Lizaola},
  {Long}, {Lucatello}, {Lundgren}, {MacDonald}, {Deconto Machado}, {MacLeod},
  {Mahadevan}, {Geimba Maia}, {Maiolino}, {Majewski}, {Malanushenko},
  {Malanushenko}, {Manchado}, {Mao}, {Maraston}, {Marques-Chaves}, {Masseron},
  {Masters}, {McBride}, {McDermid}, {McGrath}, {McGreer}, {Medina Pe{\~n}a},
  {Melendez}, {Merloni}, {Merrifield}, {Meszaros}, {Meza}, {Minchev},
  {Minniti}, {Miyaji}, {More}, {Mulchaey}, {M{\"u}ller-S{\'a}nchez}, {Muna},
  {Munoz}, {Myers}, {Nair}, {Nandra}, {Correa do Nascimento}, {Negrete},
  {Ness}, {Newman}, {Nichol}, {Nidever}, {Nitschelm}, {Ntelis}, {O'Connell},
  {Oelkers}, {Oravetz}, {Oravetz}, {Pace}, {Padilla}, {Palanque-Delabrouille},
  {Alonso Palicio}, {Pan}, {Parejko}, {Parikh}, {P{\^a}ris}, {Park}, {Patten},
  {Peirani}, {Pellejero-Ibanez}, {Penny}, {Percival}, {Perez-Fournon},
  {Petitjean}, {Pieri}, {Pinsonneault}, {Pisani}, {Poleski}, {Prada},
  {Prakash}, {Queiroz}, {Raddick}, {Raichoor}, {Barboza Rembold}, {Richstein},
  {Riffel}, {Riffel}, {Rix}, {Robin}, {Rockosi}, {Rodr{\'\i}guez-Torres},
  {Roman-Lopes}, {Rom{\'a}n-Z{\'u}{\~n}iga}, {Rosado}, {Ross}, {Rossi}, {Ruan},
  {Ruggeri}, {Rykoff}, {Salazar-Albornoz}, {Salvato}, {S{\'a}nchez}, {Aguado},
  {S{\'a}nchez-Gallego}, {Santana}, {Santiago}, {Sayres}, {Schiavon}, {da Silva
  Schimoia}, {Schlafly}, {Schlegel}, {Schneider}, {Schultheis}, {Schuster},
  {Schwope}, {Seo}, {Shao}, {Shen}, {Shetrone}, {Shull}, {Simon}, {Skinner},
  {Skrutskie}, {Slosar}, {Smith}, {Sobeck}, {Sobreira}, {Somers}, {Souto},
  {Stark}, {Stassun}, {Stauffer}, {Steinmetz}, {Storchi-Bergmann},
  {Streblyanska}, {Stringfellow}, {Su{\'a}rez}, {Sun}, {Suzuki}, {Szigeti},
  {Taghizadeh-Popp}, {Tang}, {Tao}, {Tayar}, {Tembe}, {Teske}, {Thakar},
  {Thomas}, {Thompson}, {Tinker}, {Tissera}, {Tojeiro}, {Hernandez Toledo}, {de
  la Torre}, {Tremonti}, {Troup}, {Valenzuela}, {Martinez Valpuesta},
  {Vargas-Gonz{\'a}lez}, {Vargas-Maga{\~n}a}, {Vazquez}, {Villanova}, {Vivek},
  {Vogt}, {Wake}, {Walterbos}, {Wang}, {Weaver}, {Weijmans}, {Weinberg},
  {Westfall}, {Whelan}, {Wild}, {Wilson}, {Wood-Vasey}, {Wylezalek}, {Xiao},
  {Yan}, {Yang}, {Ybarra}, {Y{\`e}che}, {Zakamska}, {Zamora}, {Zarrouk},
  {Zasowski}, {Zhang}, {Zhao}, {Zheng}, {Zheng}, {Zhou}, {Zhou}, {Zhu},
  {Zoccali}, \& {Zou}}]{sdss4}
{Blanton}, M.~R., {Bershady}, M.~A., {Abolfathi}, B., {et~al.} 2017, \aj, 154,
  28, \dodoi{10.3847/1538-3881/aa7567}

\bibitem[{Bohlin {et~al.}(2017)Bohlin, Mészáros, Fleming, Gordon, Koekemoer,
  \& Kovács}]{BOSZ}
Bohlin, R.~C., Mészáros, S., Fleming, S.~W., {et~al.} 2017, The Astronomical
  Journal, 153, 234, \dodoi{10.3847/1538-3881/aa6ba9}

\bibitem[{Bressan {et~al.}(2012)Bressan, Marigo, Girardi, Salasnich, Dal~Cero,
  Rubele, \& Nanni}]{Bressan_2012}
Bressan, A., Marigo, P., Girardi, L., {et~al.} 2012, Monthly Notices of the
  Royal Astronomical Society, 427, 127–145,
  \dodoi{10.1111/j.1365-2966.2012.21948.x}

\bibitem[{Brewer {et~al.}(2016)Brewer, Fischer, Valenti, \&
  Piskunov}]{Brewer2016}
Brewer, J.~M., Fischer, D.~A., Valenti, J.~A., \& Piskunov, N. 2016, The
  Astrophysical Journal Supplement Series, 225, 32,
  \dodoi{10.3847/0067-0049/225/2/32}

\bibitem[{{Bruzual} \& {Charlot}(2003)}]{Bruzual_2003}
{Bruzual}, G., \& {Charlot}, S. 2003, \mnras, 344, 1000,
  \dodoi{10.1046/j.1365-8711.2003.06897.x}

\bibitem[{{Bruzual} \& {Charlot}(2011)}]{Bruzual2011}
---. 2011, {GALAXEV: Evolutionary Stellar Population Synthesis Models}.
\newblock \doeprint{1104.005}

\bibitem[{{Bruzual A.}(1983)}]{Bruzual1983}
{Bruzual A.}, G. 1983, \apj, 273, 105, \dodoi{10.1086/161352}

\bibitem[{{Bundy} {et~al.}(2015){Bundy}, {Bershady}, {Law}, {Yan}, {Drory},
  {MacDonald}, {Wake}, {Cherinka}, {S{\'a}nchez-Gallego}, {Weijmans}, {Thomas},
  {Tremonti}, {Masters}, {Coccato}, {Diamond-Stanic}, {Arag{\'o}n-Salamanca},
  {Avila-Reese}, {Badenes}, {Falc{\'o}n-Barroso}, {Belfiore}, {Bizyaev},
  {Blanc}, {Bland-Hawthorn}, {Blanton}, {Brownstein}, {Byler}, {Cappellari},
  {Conroy}, {Dutton}, {Emsellem}, {Etherington}, {Frinchaboy}, {Fu}, {Gunn},
  {Harding}, {Johnston}, {Kauffmann}, {Kinemuchi}, {Klaene}, {Knapen},
  {Leauthaud}, {Li}, {Lin}, {Maiolino}, {Malanushenko}, {Malanushenko}, {Mao},
  {Maraston}, {McDermid}, {Merrifield}, {Nichol}, {Oravetz}, {Pan}, {Parejko},
  {Sanchez}, {Schlegel}, {Simmons}, {Steele}, {Steinmetz}, {Thanjavur},
  {Thompson}, {Tinker}, {van den Bosch}, {Westfall}, {Wilkinson}, {Wright},
  {Xiao}, \& {Zhang}}]{MaNGAOverview}
{Bundy}, K., {Bershady}, M.~A., {Law}, D.~R., {et~al.} 2015, \apj, 798, 7,
  \dodoi{10.1088/0004-637X/798/1/7}

\bibitem[{{Cardiel} {et~al.}(2003){Cardiel}, {Gorgas},
  {S{\'a}nchez-Bl{\'a}zquez}, {Cenarro}, {Pedraz}, {Bruzual}, \&
  {Klement}}]{Cardiel2003}
{Cardiel}, N., {Gorgas}, J., {S{\'a}nchez-Bl{\'a}zquez}, P., {et~al.} 2003,
  \aap, 409, 511, \dodoi{10.1051/0004-6361:20031096}

\bibitem[{Casey {et~al.}(2016)Casey, Hogg, Ness, Rix, Ho, \& Gilmore}]{cannon2}
Casey, A.~R., Hogg, D.~W., Ness, M., {et~al.} 2016, The Cannon 2: A data-driven
  model of stellar spectra for detailed chemical abundance analyses.
\newblock \doarXiv{1603.03040}

\bibitem[{{Castelli} \& {Kurucz}(2003)}]{atlas9}
{Castelli}, F., \& {Kurucz}, R.~L. 2003, in Modelling of Stellar Atmospheres,
  ed. N.~{Piskunov}, W.~W. {Weiss}, \& D.~F. {Gray}, Vol. 210, A20.
\newblock \doarXiv{astro-ph/0405087}

\bibitem[{{Cenarro} {et~al.}(2001){Cenarro}, {Cardiel}, {Gorgas}, {Peletier},
  {Vazdekis}, \& {Prada}}]{Cenarro2001}
{Cenarro}, A.~J., {Cardiel}, N., {Gorgas}, J., {et~al.} 2001, \mnras, 326, 959,
  \dodoi{10.1046/j.1365-8711.2001.04688.x}

\bibitem[{Chen {et~al.}(2021 in prep.)Chen, Yan, \& et~al.}]{Chen2021}
Chen, Y., Yan, R., \& et~al. 2021 in prep.

\bibitem[{{Chen} {et~al.}(2014){Chen}, {Trager}, {Peletier}, {Lan{\c{c}}on},
  {Vazdekis}, {Prugniel}, {Silva}, \& {Gonneau}}]{Chen2014}
{Chen}, Y.-P., {Trager}, S.~C., {Peletier}, R.~F., {et~al.} 2014, \aap, 565,
  A117, \dodoi{10.1051/0004-6361/201322505}

\bibitem[{Chen {et~al.}(2020)Chen, Yan, Maraston, Thomas, Stringfellow,
  Bizyaev, Gelfand, Beers, Fernández-Trincado, Lazarz, \& et~al.}]{Chen_2020}
Chen, Y.-P., Yan, R., Maraston, C., {et~al.} 2020, The Astrophysical Journal,
  899, 62, \dodoi{10.3847/1538-4357/ab9f35}

\bibitem[{{Coelho} {et~al.}(2007){Coelho}, {Bruzual}, {Charlot}, {Weiss},
  {Barbuy}, \& {Ferguson}}]{Coelho2007}
{Coelho}, P., {Bruzual}, G., {Charlot}, S., {et~al.} 2007, \mnras, 382, 498,
  \dodoi{10.1111/j.1365-2966.2007.12364.x}

\bibitem[{Conroy(2013)}]{Conroy_2013}
Conroy, C. 2013, Annual Review of Astronomy and Astrophysics, 51, 393–455,
  \dodoi{10.1146/annurev-astro-082812-141017}

\bibitem[{Conroy {et~al.}(2009)Conroy, Gunn, \& White}]{Conroy_2009}
Conroy, C., Gunn, J.~E., \& White, M. 2009, The Astrophysical Journal, 699,
  486–506, \dodoi{10.1088/0004-637x/699/1/486}

\bibitem[{{Croom} {et~al.}(2021){Croom}, {Owers}, {Scott}, {Poetrodjojo},
  {Groves}, {van de Sande}, {Barone}, {Cortese}, {D'Eugenio}, {Bland-Hawthorn},
  {Bryant}, {Oh}, {Brough}, {Agostino}, {Casura}, {Catinella}, {Colless},
  {Cecil}, {Davies}, {Drinkwater}, {Driver}, {Ferreras}, {Foster},
  {Fraser-McKelvie}, {Lawrence}, {Leslie}, {Liske}, {L{\'o}pez-S{\'a}nchez},
  {Lorente}, {McElroy}, {Medling}, {Obreschkow}, {Richards}, {Sharp}, {Sweet},
  {Taranu}, {Taylor}, {Tescari}, {Thomas}, {Tocknell}, \&
  {Vaughan}}]{Croom2021}
{Croom}, S.~M., {Owers}, M.~S., {Scott}, N., {et~al.} 2021, Monthly Notices of
  the Royal Astronomical Society, \dodoi{10.1093/mnras/stab229}

\bibitem[{da~Silva {et~al.}(2015)da~Silva, Milone, \&
  Rocha-Pinto}]{daSilva2015}
da~Silva, R., Milone, A. d.~C., \& Rocha-Pinto, H.~J. 2015, Astronomy \&
  Astrophysics, 580, A24, \dodoi{10.1051/0004-6361/201525770}

\bibitem[{Decin {et~al.}(2004)Decin, Morris, Appleton, Charmandaris, Armus, \&
  Houck}]{Decin2004}
Decin, L., Morris, P.~W., Appleton, P.~N., {et~al.} 2004, The Astrophysical
  Journal Supplement Series, 154, 408–412, \dodoi{10.1086/422884}

\bibitem[{{Diaz} {et~al.}(1989){Diaz}, {Terlevich}, \& {Terlevich}}]{Diaz1989}
{Diaz}, A.~I., {Terlevich}, E., \& {Terlevich}, R. 1989, \mnras, 239, 325,
  \dodoi{10.1093/mnras/239.2.325}

\bibitem[{{Drory} {et~al.}(2015){Drory}, {MacDonald}, {Bershady}, {Bundy},
  {Gunn}, {Law}, {Smith}, {Stoll}, {Tremonti}, {Wake}, {Yan}, {Weijmans},
  {Byler}, {Cherinka}, {Cope}, {Eigenbrot}, {Harding}, {Holder}, {Huehnerhoff},
  {Jaehnig}, {Jansen}, {Klaene}, {Paat}, {Percival}, \& {Sayres}}]{Drory2015}
{Drory}, N., {MacDonald}, N., {Bershady}, M.~A., {et~al.} 2015, \aj, 149, 77,
  \dodoi{10.1088/0004-6256/149/2/77}

\bibitem[{Dupree {et~al.}(2016)Dupree, Avrett, \& Kurucz}]{Dupree_2016}
Dupree, A.~K., Avrett, E.~H., \& Kurucz, R.~L. 2016, The Astrophysical Journal,
  821, L7, \dodoi{10.3847/2041-8205/821/1/l7}

\bibitem[{{Fioc} \& {Rocca-Volmerange}(1997)}]{Fioc1997}
{Fioc}, M., \& {Rocca-Volmerange}, B. 1997, \aap, 500, 507.
\newblock \doarXiv{astro-ph/9707017}

\bibitem[{{Forsberg} {et~al.}(2019){Forsberg}, {J{\"o}nsson}, {Ryde}, \&
  {Matteucci}}]{Forseberg2019}
{Forsberg}, R., {J{\"o}nsson}, H., {Ryde}, N., \& {Matteucci}, F. 2019, \aap,
  631, A113, \dodoi{10.1051/0004-6361/201936343}

\bibitem[{{Fuhrmann}(1998)}]{Fuhrmann1998}
{Fuhrmann}, K. 1998, \aap, 338, 161

\bibitem[{García-Benito {et~al.}(2017)García-Benito, González~Delgado,
  Pérez, Cid~Fernandes, Cortijo-Ferrero, López~Fernández, de~Amorim,
  Lacerda, Vale~Asari, Sánchez, \& et~al.}]{GarciaBenito2017}
García-Benito, R., González~Delgado, R.~M., Pérez, E., {et~al.} 2017,
  Astronomy \& Astrophysics, 608, A27, \dodoi{10.1051/0004-6361/201731357}

\bibitem[{García~Pérez {et~al.}(2016)García~Pérez, Prieto, Holtzman,
  Shetrone, Mészáros, Bizyaev, Carrera, Cunha, García-Hernández, Johnson,
  \& et~al.}]{aspcap}
García~Pérez, A.~E., Prieto, C.~A., Holtzman, J.~A., {et~al.} 2016, The
  Astronomical Journal, 151, 144, \dodoi{10.3847/0004-6256/151/6/144}

\bibitem[{Goddard {et~al.}(2016)Goddard, Thomas, Maraston, Westfall,
  Etherington, Riffel, Mallmann, Zheng, Argudo-Fernández, Lian, \&
  et~al.}]{Goddard2016}
Goddard, D., Thomas, D., Maraston, C., {et~al.} 2016, Monthly Notices of the
  Royal Astronomical Society, stw3371, \dodoi{10.1093/mnras/stw3371}

\bibitem[{{Gonz{\'a}lez Hern{\'a}ndez} \& {Bonifacio}(2009)}]{GHB2009}
{Gonz{\'a}lez Hern{\'a}ndez}, J.~I., \& {Bonifacio}, P. 2009, \aap, 497, 497,
  \dodoi{10.1051/0004-6361/200810904}

\bibitem[{{Gregg} {et~al.}(2006){Gregg}, {Silva}, {Rayner}, {Worthey},
  {Valdes}, {Pickles}, {Rose}, {Carney}, \& {Vacca}}]{gregg2006}
{Gregg}, M.~D., {Silva}, D., {Rayner}, J., {et~al.} 2006, in The 2005 HST
  Calibration Workshop: Hubble After the Transition to Two-Gyro Mode, ed. A.~M.
  {Koekemoer}, P.~{Goudfrooij}, \& L.~L. {Dressel}, 209

\bibitem[{{Gunn} \& {Stryker}(1983)}]{Gunn1983}
{Gunn}, J.~E., \& {Stryker}, L.~L. 1983, \apjs, 52, 121, \dodoi{10.1086/190861}

\bibitem[{{Gunn} {et~al.}(2006){Gunn}, {Siegmund}, {Mannery}, {Owen}, {Hull},
  {Leger}, {Carey}, {Knapp}, {York}, {Boroski}, {Kent}, {Lupton}, {Rockosi},
  {Evans}, {Waddell}, {Anderson}, {Annis}, {Barentine}, {Bartoszek}, {Bastian},
  {Bracker}, {Brewington}, {Briegel}, {Brinkmann}, {Brown}, {Carr},
  {Czarapata}, {Drennan}, {Dombeck}, {Federwitz}, {Gillespie}, {Gonzales},
  {Hansen}, {Harvanek}, {Hayes}, {Jordan}, {Kinney}, {Klaene}, {Kleinman},
  {Kron}, {Kresinski}, {Lee}, {Limmongkol}, {Lindenmeyer}, {Long}, {Loomis},
  {McGehee}, {Mantsch}, {Neilsen}, {Neswold}, {Newman}, {Nitta}, {Peoples},
  {Pier}, {Prieto}, {Prosapio}, {Rivetta}, {Schneider}, {Snedden}, \&
  {Wang}}]{apo25m}
{Gunn}, J.~E., {Siegmund}, W.~A., {Mannery}, E.~J., {et~al.} 2006, \aj, 131,
  2332, \dodoi{10.1086/500975}

\bibitem[{{Gustafsson} {et~al.}(2008){Gustafsson}, {Edvardsson}, {Eriksson},
  {J{\o}rgensen}, {Nordlund}, \& {Plez}}]{MARCS2008}
{Gustafsson}, B., {Edvardsson}, B., {Eriksson}, K., {et~al.} 2008, \aap, 486,
  951, \dodoi{10.1051/0004-6361:200809724}

\bibitem[{Hill {et~al.}(2021 submitted)Hill, Thomas, Maraston, Yan, \&
  et~al.}]{hill2021}
Hill, L., Thomas, D., Maraston, C., Yan, R., \& et~al. 2021 submitted, SDSS-IV
  MaStar: Fundamental Atmospheric Parameters for the MaNGA Stellar Library

\bibitem[{{Holtzman et al.}(2021 in prep.)}]{aspcapdr17}
{Holtzman et al.}, J.~A. 2021 in prep.

\bibitem[{{J{\"o}nsson} {et~al.}(2017){J{\"o}nsson}, {Ryde}, {Nordlander},
  {Pehlivan Rhodin}, {Hartman}, {J{\"o}nsson}, \& {Eriksson}}]{Jonsson2017}
{J{\"o}nsson}, H., {Ryde}, N., {Nordlander}, T., {et~al.} 2017, \aap, 598,
  A100, \dodoi{10.1051/0004-6361/201629128}

\bibitem[{Jönsson {et~al.}(2018)Jönsson, Prieto, Holtzman, Feuillet, Hawkins,
  Cunha, Mészáros, Hasselquist, Fernández-Trincado, García-Hernández, \&
  et~al.}]{aspcap2018}
Jönsson, H., Prieto, C.~A., Holtzman, J.~A., {et~al.} 2018, The Astronomical
  Journal, 156, 126, \dodoi{10.3847/1538-3881/aad4f5}

\bibitem[{{Kirby}(2011)}]{Kirby2011}
{Kirby}, E.~N. 2011, \pasp, 123, 531, \dodoi{10.1086/660019}

\bibitem[{Koesterke(2009)}]{koesterke2009}
Koesterke, L. 2009, AIP Conference Proceedings, 1171, 73,
  \dodoi{10.1063/1.3250090}

\bibitem[{{Koleva, M.} {et~al.}(2009){Koleva, M.}, {Prugniel, Ph.}, {Bouchard,
  A.}, \& {Wu, Y.}}]{Koleva2009}
{Koleva, M.}, {Prugniel, Ph.}, {Bouchard, A.}, \& {Wu, Y.} 2009, A\&A, 501,
  1269, \dodoi{10.1051/0004-6361/200811467}

\bibitem[{{Kurucz}(1993)}]{Kurucz1993}
{Kurucz}, R. 1993, ATLAS9 Stellar Atmosphere Programs and 2 km/s grid. Kurucz
  CD-ROM No. 13. Cambridge, 13

\bibitem[{{Kurucz}(1979)}]{Kurucz1979}
{Kurucz}, R.~L. 1979, \apjs, 40, 1, \dodoi{10.1086/190589}

\bibitem[{{Kurucz}(2011)}]{Kurucz2011}
---. 2011, Canadian Journal of Physics, 89, 417, \dodoi{10.1139/p10-104}

\bibitem[{Law {et~al.}(2016)Law, Cherinka, Yan, Andrews, Bershady, Bizyaev,
  Blanc, Blanton, Bolton, Brownstein, \& et~al.}]{mangadrp}
Law, D.~R., Cherinka, B., Yan, R., {et~al.} 2016, The Astronomical Journal,
  152, 83, \dodoi{10.3847/0004-6256/152/4/83}

\bibitem[{Lazarz {et~al.}(2021 in prep.)Lazarz, Yan, \& et~al.}]{lazarz2021}
Lazarz, D., Yan, R., \& et~al. 2021 in prep.

\bibitem[{{Le Borgne} {et~al.}(2003){Le Borgne}, {Bruzual}, {Pell{\'o}},
  {Lan{\c{c}}on}, {Rocca-Volmerange}, {Sanahuja}, {Schaerer}, {Soubiran}, \&
  {V{\'\i}lchez-G{\'o}mez}}]{LeBorgne2003}
{Le Borgne}, J.~F., {Bruzual}, G., {Pell{\'o}}, R., {et~al.} 2003, \aap, 402,
  433, \dodoi{10.1051/0004-6361:20030243}

\bibitem[{Lee {et~al.}(2008)Lee, Beers, Sivarani, Allende~Prieto, Koesterke,
  Wilhelm, Fiorentin, Bailer-Jones, Norris, Rockosi, \& et~al.}]{Lee_2008}
Lee, Y.~S., Beers, T.~C., Sivarani, T., {et~al.} 2008, The Astronomical
  Journal, 136, 2022–2049, \dodoi{10.1088/0004-6256/136/5/2022}

\bibitem[{Leitherer {et~al.}(1999)Leitherer, Schaerer, Goldader, Delgado,
  Robert, Kune, de~Mello, Devost, \& Heckman}]{Leitherer_1999}
Leitherer, C., Schaerer, D., Goldader, J.~D., {et~al.} 1999, The Astrophysical
  Journal Supplement Series, 123, 3–40, \dodoi{10.1086/313233}

\bibitem[{{Lejeune} {et~al.}(1997){Lejeune}, {Cuisinier}, \&
  {Buser}}]{Lejeune1997}
{Lejeune}, T., {Cuisinier}, F., \& {Buser}, R. 1997, \aaps, 125, 229,
  \dodoi{10.1051/aas:1997373}

\bibitem[{{Lomaeva} {et~al.}(2019){Lomaeva}, {J{\"o}nsson}, {Ryde},
  {Schultheis}, \& {Thorsbro}}]{Lomaeva2019}
{Lomaeva}, M., {J{\"o}nsson}, H., {Ryde}, N., {Schultheis}, M., \& {Thorsbro},
  B. 2019, \aap, 625, A141, \dodoi{10.1051/0004-6361/201834247}

\bibitem[{Luo {et~al.}(2015)Luo, Zhao, Zhao, Deng, Liu, Jing, Wang, Zhang, Shi,
  Cui, \& et~al.}]{LAMOST_Luo2015}
Luo, A.-L., Zhao, Y.-H., Zhao, G., {et~al.} 2015, Research in Astronomy and
  Astrophysics, 15, 1095–1124, \dodoi{10.1088/1674-4527/15/8/002}

\bibitem[{{Majewski} {et~al.}(2017){Majewski}, {Schiavon}, {Frinchaboy},
  {Allende Prieto}, {Barkhouser}, {Bizyaev}, {Blank}, {Brunner}, {Burton},
  {Carrera}, {Chojnowski}, {Cunha}, {Epstein}, {Fitzgerald}, {Garc{\'\i}a
  P{\'e}rez}, {Hearty}, {Henderson}, {Holtzman}, {Johnson}, {Lam}, {Lawler},
  {Maseman}, {M{\'e}sz{\'a}ros}, {Nelson}, {Nguyen}, {Nidever}, {Pinsonneault},
  {Shetrone}, {Smee}, {Smith}, {Stolberg}, {Skrutskie}, {Walker}, {Wilson},
  {Zasowski}, {Anders}, {Basu}, {Beland}, {Blanton}, {Bovy}, {Brownstein},
  {Carlberg}, {Chaplin}, {Chiappini}, {Eisenstein}, {Elsworth}, {Feuillet},
  {Fleming}, {Galbraith-Frew}, {Garc{\'\i}a}, {Garc{\'\i}a-Hern{\'a}ndez},
  {Gillespie}, {Girardi}, {Gunn}, {Hasselquist}, {Hayden}, {Hekker}, {Ivans},
  {Kinemuchi}, {Klaene}, {Mahadevan}, {Mathur}, {Mosser}, {Muna}, {Munn},
  {Nichol}, {O'Connell}, {Parejko}, {Robin}, {Rocha-Pinto}, {Schultheis},
  {Serenelli}, {Shane}, {Silva Aguirre}, {Sobeck}, {Thompson}, {Troup},
  {Weinberg}, \& {Zamora}}]{apogeeoverview}
{Majewski}, S.~R., {Schiavon}, R.~P., {Frinchaboy}, P.~M., {et~al.} 2017, \aj,
  154, 94, \dodoi{10.3847/1538-3881/aa784d}

\bibitem[{Maraston(2005)}]{Maraston_2005}
Maraston, C. 2005, Monthly Notices of the Royal Astronomical Society, 362,
  799–825, \dodoi{10.1111/j.1365-2966.2005.09270.x}

\bibitem[{Maraston \& Strömbäck(2011)}]{Maraston_2011}
Maraston, C., \& Strömbäck, G. 2011, Monthly Notices of the Royal
  Astronomical Society, 418, 2785, \dodoi{10.1111/j.1365-2966.2011.19738.x}

\bibitem[{{Maraston} {et~al.}(2020){Maraston}, {Hill}, {Thomas}, {Yan}, {Chen},
  {Lian}, {Parikh}, {Neumann}, {Meneses-Goytia}, {Bershady}, {Drory},
  {Bizyaev}, {Concas}, {Brownstein}, {Lazarz}, {Stringfellow}, \&
  {Stassun}}]{maraston2020}
{Maraston}, C., {Hill}, L., {Thomas}, D., {et~al.} 2020, \mnras, 496, 2962,
  \dodoi{10.1093/mnras/staa1489}

\bibitem[{{Matteucci} \& {Brocato}(1990)}]{mb1990}
{Matteucci}, F., \& {Brocato}, E. 1990, \apj, 365, 539, \dodoi{10.1086/169508}

\bibitem[{{M{\'e}sz{\'a}ros} {et~al.}(2012){M{\'e}sz{\'a}ros}, {Allende
  Prieto}, {Edvardsson}, {Castelli}, {Garc{\'\i}a P{\'e}rez}, {Gustafsson},
  {Majewski}, {Plez}, {Schiavon}, {Shetrone}, \& {de Vicente}}]{apogeegrid}
{M{\'e}sz{\'a}ros}, S., {Allende Prieto}, C., {Edvardsson}, B., {et~al.} 2012,
  \aj, 144, 120, \dodoi{10.1088/0004-6256/144/4/120}

\bibitem[{Montalbán {et~al.}(2007)Montalbán, Nendwich, Heiter, Kupka,
  Paunzen, \& Smalley}]{Montalban2007}
Montalbán, J., Nendwich, J., Heiter, U., {et~al.} 2007, Proceedings of The
  International Astronomical Union, 239, 166, \dodoi{10.1017/S1743921307000361}

\bibitem[{Ness {et~al.}(2015)Ness, Hogg, Rix, Ho, \& Zasowski}]{cannon1}
Ness, M., Hogg, D.~W., Rix, H.-W., Ho, A. Y.~Q., \& Zasowski, G. 2015, The
  Astrophysical Journal, 808, 16, \dodoi{10.1088/0004-637x/808/1/16}

\bibitem[{Percival {et~al.}(2008)Percival, Salaris, Cassisi, \&
  Pietrinferni}]{Percival2008}
Percival, S.~M., Salaris, M., Cassisi, S., \& Pietrinferni, A. 2008, The
  Astrophysical Journal, 690, 427–439, \dodoi{10.1088/0004-637x/690/1/427}

\bibitem[{{Pickles}(1985)}]{Pickles1985}
{Pickles}, A.~J. 1985, \apjs, 59, 33, \dodoi{10.1086/191061}

\bibitem[{{Pickles}(1998)}]{Pickles1998}
---. 1998, \pasp, 110, 863, \dodoi{10.1086/316197}

\bibitem[{{Prugniel} \& {Soubiran}(2001)}]{Prugniel2001}
{Prugniel}, P., \& {Soubiran}, C. 2001, \aap, 369, 1048,
  \dodoi{10.1051/0004-6361:20010163}

\bibitem[{Rix {et~al.}(2016)Rix, Ting, Conroy, \& Hogg}]{Rix2016}
Rix, H.-W., Ting, Y.-S., Conroy, C., \& Hogg, D.~W. 2016, The Astrophysical
  Journal, 826, L25, \dodoi{10.3847/2041-8205/826/2/l25}

\bibitem[{Röck {et~al.}(2016)Röck, Vazdekis, Ricciardelli, Peletier, Knapen,
  \& Falcón-Barroso}]{R_ck_2016}
Röck, B., Vazdekis, A., Ricciardelli, E., {et~al.} 2016, Astronomy \&
  Astrophysics, 589, A73, \dodoi{10.1051/0004-6361/201527570}

\bibitem[{Sanchez-Blazquez {et~al.}(2006)Sanchez-Blazquez, Peletier,
  Jimenez-Vicente, Cardiel, Cenarro, Falcon-Barroso, Gorgas, Selam, \&
  Vazdekis}]{MILES2006}
Sanchez-Blazquez, P., Peletier, R.~F., Jimenez-Vicente, J., {et~al.} 2006,
  Monthly Notices of the Royal Astronomical Society, 371, 703–718,
  \dodoi{10.1111/j.1365-2966.2006.10699.x}

\bibitem[{{Silva} \& {Cornell}(1992)}]{Silva1992}
{Silva}, D.~R., \& {Cornell}, M.~E. 1992, \apjs, 81, 865,
  \dodoi{10.1086/191706}

\bibitem[{{Smee} {et~al.}(2013){Smee}, {Gunn}, {Uomoto}, {Roe}, {Schlegel},
  {Rockosi}, {Carr}, {Leger}, {Dawson}, {Olmstead}, {Brinkmann}, {Owen},
  {Barkhouser}, {Honscheid}, {Harding}, {Long}, {Lupton}, {Loomis}, {Anderson},
  {Annis}, {Bernardi}, {Bhardwaj}, {Bizyaev}, {Bolton}, {Brewington}, {Briggs},
  {Burles}, {Burns}, {Castander}, {Connolly}, {Davenport}, {Ebelke}, {Epps},
  {Feldman}, {Friedman}, {Frieman}, {Heckman}, {Hull}, {Knapp}, {Lawrence},
  {Loveday}, {Mannery}, {Malanushenko}, {Malanushenko}, {Merrelli}, {Muna},
  {Newman}, {Nichol}, {Oravetz}, {Pan}, {Pope}, {Ricketts}, {Shelden},
  {Sandford}, {Siegmund}, {Simmons}, {Smith}, {Snedden}, {Schneider},
  {SubbaRao}, {Tremonti}, {Waddell}, \& {York}}]{bossspectrograph}
{Smee}, S.~A., {Gunn}, J.~E., {Uomoto}, A., {et~al.} 2013, \aj, 146, 32,
  \dodoi{10.1088/0004-6256/146/2/32}

\bibitem[{{Smith} {et~al.}(2021){Smith}, {Bizyaev}, {Cunha}, {Shetrone},
  {Souto}, {Allende Prieto}, {Masseron}, {M{\'e}sz{\'a}ros}, {J{\"o}nsson},
  {Hasselquist}, {Osorio}, {Garc{\'\i}a-Hern{\'a}ndez}, {Plez}, {Beaton},
  {Holtzman}, {Majewski}, {Stringfellow}, \& {Sobeck}}]{Smith2021}
{Smith}, V.~V., {Bizyaev}, D., {Cunha}, K., {et~al.} 2021, \aj, 161, 254,
  \dodoi{10.3847/1538-3881/abefdc}

\bibitem[{Sánchez {et~al.}(2016)Sánchez, García-Benito, Zibetti, Walcher,
  Husemann, Mendoza, Galbany, Falcón-Barroso, Mast, Aceituno, \&
  et~al.}]{CALIFA}
Sánchez, S.~F., García-Benito, R., Zibetti, S., {et~al.} 2016, Astronomy \&
  Astrophysics, 594, A36, \dodoi{10.1051/0004-6361/201628661}

\bibitem[{{Thomas} {et~al.}(2005){Thomas}, {Maraston}, {Bender}, \& {Mendes de
  Oliveira}}]{Thomas2005}
{Thomas}, D., {Maraston}, C., {Bender}, R., \& {Mendes de Oliveira}, C. 2005,
  \apj, 621, 673, \dodoi{10.1086/426932}

\bibitem[{Ting {et~al.}(2019)Ting, Conroy, Rix, \& Cargile}]{Ting_2019}
Ting, Y.-S., Conroy, C., Rix, H.-W., \& Cargile, P. 2019, The Astrophysical
  Journal, 879, 69, \dodoi{10.3847/1538-4357/ab2331}

\bibitem[{{Tinsley}(1972)}]{Tinsley1972}
{Tinsley}, B.~M. 1972, \apj, 178, 319, \dodoi{10.1086/151793}

\bibitem[{Valdes {et~al.}(2004)Valdes, Gupta, Rose, Singh, \&
  Bell}]{Valdes2004}
Valdes, F., Gupta, R., Rose, J.~A., Singh, H.~P., \& Bell, D.~J. 2004, The
  Astrophysical Journal Supplement Series, 152, 251–259,
  \dodoi{10.1086/386343}

\bibitem[{Vazdekis {et~al.}(2016)Vazdekis, Koleva, Ricciardelli, Röck, \&
  Falcón-Barroso}]{Vazdekis2016}
Vazdekis, A., Koleva, M., Ricciardelli, E., Röck, B., \& Falcón-Barroso, J.
  2016, Monthly Notices of the Royal Astronomical Society, 463, 3409–3436,
  \dodoi{10.1093/mnras/stw2231}

\bibitem[{Vazdekis {et~al.}(2010)Vazdekis, Sánchez-Blázquez, Falcón-Barroso,
  Cenarro, Beasley, Cardiel, Gorgas, \& Peletier}]{Vazdekis_2010}
Vazdekis, A., Sánchez-Blázquez, P., Falcón-Barroso, J., {et~al.} 2010,
  Monthly Notices of the Royal Astronomical Society,
  \dodoi{10.1111/j.1365-2966.2010.16407.x}

\bibitem[{{Westera} {et~al.}(2002){Westera}, {Lejeune}, {Buser}, {Cuisinier},
  \& {Bruzual}}]{Westera2002}
{Westera}, P., {Lejeune}, T., {Buser}, R., {Cuisinier}, F., \& {Bruzual}, G.
  2002, \aap, 381, 524, \dodoi{10.1051/0004-6361:20011493}

\bibitem[{Yan {et~al.}(2021 in prep.)Yan, Chen, Hill, Imig, Lazarz, \&
  et~al.}]{mastar2021}
Yan, R., Chen, Y.-P., Hill, L., {et~al.} 2021 in prep.

\bibitem[{Yan {et~al.}(2019)Yan, Chen, Lazarz, Bizyaev, Maraston, Stringfellow,
  McCarthy, Meneses-Goytia, Law, Thomas, \& et~al.}]{mastar2019}
Yan, R., Chen, Y., Lazarz, D., {et~al.} 2019, The Astrophysical Journal, 883,
  175, \dodoi{10.3847/1538-4357/ab3ebc}

\end{thebibliography}
\bibliographystyle{aasjournal}



\end{document}